\documentclass[12pt,preprint]{aastex}

\usepackage{bm}
\shorttitle{On the Sensitivity of PRD Scattering  Polarization Profiles to Various Atmospheric Parameters}
\shortauthors{Sampoorna et al.}
\begin{document}
\title{On the Sensitivity of Partial Redistribution Scattering Polarization 
Profiles to Various Atmospheric Parameters 
}
\author{M. Sampoorna$^{1,}$\altaffilmark{\dag}, J. Trujillo Bueno$^{1,2,3}$, and E. Landi Degl'Innocenti$^{1,4}$} 
\affil{$^1$Instituto de Astrof\'isica de Canarias, E-38205 La Laguna, Tenerife, Spain}
\affil{$^2$Departamento de Astrof\'isica, Facultad de F\'isica, Universidad de 
La Laguna, \\ Tenerife, Spain}
\affil{$^3$Consejo Superior de Investigaciones Cient\'ificas, Spain}
\affil{$^4$Dipartimento di Fisica e Astronomia, Sezione di Astronomia  
e Scienza dello Spazio, \\ Universit\`a degli Studi di Firenze, 
Largo Enrico Fermi 2, I-50125 Firenze, Italy}
\email{sampoorna@iiap.res.in; jtb@iac.es; landie@arcetri.astro.it}
\affil{{\bf{Accepted in August 2010 for publication in The Astrophysical Journal}}}
\altaffiltext{\dag}{Presently at Indian Institute of Astrophysics, Koramangala, 
Bangalore, India}
\begin{abstract}
This paper presents a detailed study of the scattering polarization profiles 
formed under partial frequency redistribution (PRD) in two thermal models of 
the solar atmosphere. Particular attention is given to understanding the 
influence of several atmospheric parameters on the emergent fractional linear 
polarization profiles. The shapes of these $Q/I$ profiles are interpreted 
in terms of the anisotropy 
of the radiation field, which in turn depends on the source function 
gradient that sets the angular variation of the specific intensity. We define 
a suitable frequency integrated anisotropy factor for PRD that can be 
directly related to the emergent linear polarization. We show that complete 
frequency redistribution is a good approximation to model weak 
resonance lines. We also 
show that the emergent linear polarization profiles can be very sensitive to the thermal 
structure of the solar atmosphere and, in particular, to spatial variations of the 
damping parameter.
\end{abstract}

\section{Introduction}
The linearly polarized solar limb spectrum 
produced by scattering processes in quiet regions of the solar atmosphere 
\citep{josck97}, contains a wealth of information on the physics of 
scattering and on the thermodynamical and magnetic conditions of the solar 
atmosphere. The rich structuring of this so-called second solar spectrum and 
its interpretation has opened a new window in solar physics  
with great diagnostic potential \citep[e.g.,][]{jos04,jos06,jtb01,jtb03,jtb09}. 
This linearly polarized spectrum has been 
measured with high spectral resolution from the UV at
3160\,\AA\ to the red at 6995\,\AA\ \citep{gan00,gan02,gan05}. 
Using this atlas of the second solar spectrum \citet{lbandeld09} have presented a 
very useful classification and spectroscopic analysis of the most polarizing 
atomic lines. 

The modeling and physical interpretation of the second solar spectrum 
requires the application of sophisticated theories of line formation 
capable of accounting for, in a self-consistent way, the physics of 
scattering polarization in the presence of magnetic fields.
Starting from the principles of quantum 
electrodynamics, a rigorous theory of the polarization in spectral lines 
was formulated by \citet{vbandsb78} for optically thin lines and by 
\citet[][see also the monograph by Landi Degl'Innocenti \& Landolfi 2004]{landi83} 
for optically thick lines. This latter theory, which can take into account 
radiative transfer effects in 
multi-level atomic systems, lower level polarization, hyperfine structure and 
level crossings interferences, has been successfully applied 
to interpret several interesting spectropolarimetric observations 
\citep[see][for a recent review]{jtb09}. However, a limitation 
of this theory is that it is based on the approximation of 
complete frequency redistribution (CRD) in scattering (namely, no 
correlation between the frequencies of the incident and scattered photon). 
Though this 
is a suitable approximation for several solar spectral lines 
\citep[e.g.,][]{ms-jtb03,jtb-nature04,stepan-jtb10},  
it is not suitable to model the observed linear polarization patterns in 
some of the strongest  and most prominent lines of the second solar spectrum, like Ca\,{\sc i} 
4227\,\AA\ and Ca\,{\sc ii} K at 3933 \AA. For these and a few other resonance 
lines one has to take into account the effects of partial frequency 
redistribution (PRD) in scattering. 

An attempt to handle PRD effects 
within a theoretical framework where the basic results of the CRD density-matrix theory are 
generalized through the hypothesis of considering the atomic levels as a 
continuous distribution of infinitely sharp sublevels (the so-called metalevels
theory) has been proposed by \citet{landietal97}. Although this theory has 
proved to be quite successful for some applications 
\citep[e.g.,][]{landi98} the difficulty of coherently 
incorporating depolarizing collisional effects in such a  theoretical approach 
prevents us from employing it in the present investigation. We thus prefer, in this 
paper, to rely on the better-established theory of resonance scattering 
(see below) although we have to remind that such a theory suffers from the 
main limitation of being applicable only for two-level atoms with unpolarized, 
infinitely sharp lower-levels.

In a scattering event, the direction, frequency, 
and polarization of the scattered photon are in general different from those 
of the incident photon. The correlations between angle, frequency and 
polarization of the incident and scattered photons can be described by a PRD 
matrix. The effects of PRD are clearly  
observable in the wings of strong resonance lines and to a lesser extent in 
the line core \citep[e.g.,][]{hf96}. The general 
problem of redistribution of resonance radiation including the effects of 
collisions was investigated by \citet{osc72}. Using a quantum mechanical description of 
matter and radiation, they derived PRD functions in the rest frame of the 
atom. A year later, these authors addressed the same problem but  
for the magnetized case \citep{osc73}. Starting from the work of 
\citet{osc72}, \citet{dh88} derived expressions of the 
PRD matrices for resonance line polarization in a two-level 
atom with unpolarized lower-level. By applying a master equation theory, 
\citet{vb97a} derived a more elegant but equivalent expression 
for the non-magnetic PRD matrix. Furthermore, she 
derived the explicit form of the laboratory frame PRD  
matrix in the presence of an arbitrary magnetic field \citep[see][]{vb97b}, 
for the case of a two-level model atom without atomic 
polarization in the lower level. \citet{bs99}
developed a classical time-dependent oscillator PRD theory 
for the particular case of a normal Zeeman triplet 
in the atomic rest frame. 
The ensuing laboratory frame PRD matrices 
for the magnetized case were derived by \citet{sametal07a,sametal07b}. 
In this paper we restrict our attention to the problem of resonance line 
polarization in situations where the radiation field is axially symmetric 
(e.g., the case of non-magnetic resonance line polarization in plane-parallel 
stellar atmospheres). We use the angle-averaged (AA) version of the 
non-magnetic PRD matrix \citep[see][]{dh88,vb97b}. 

\citet{dumontetal77} carried out model calculations of polarized line transfer 
with PRD in axi-symmetric plane-parallel atmospheres. They used the 
angle-dependent type I (pure Doppler) redistribution function of 
\citet{hum62}, which may be used in some circumstances for the Doppler 
cores of resonance lines. They considered (1) an 
isothermal model atmosphere and (2) a chromospheric-like model atmosphere 
described by an analytic form for the Planck function. For the case of 
isothermal model atmospheres they studied the influence of $\epsilon$ 
(the photon destruction probability per scattering event) and the error 
introduced by the assumption of CRD on the emergent linear polarization 
profiles. They showed that CRD is adequate to describe the line core 
polarization. For the case of schematic chromospheric-like model atmospheres, 
they considered the influence of the line strength, but under the assumption of CRD. 
\citet[][see also Saliba 1985, 1986]{rs82} considered the same problem, 
but using an AA type II redistribution function 
($R_{\rm II,AA}$) of \citet{hum62}. 
However, for $R_{\rm II,AA}$ they used an approximate form 
given by \citet{kf75}, which assumes that CRD prevails in the line 
core and coherent scattering in the wings. They studied the influence 
of the line strength, $\epsilon$ and damping parameter $a$ of the absorption 
profile on the emergent linear polarization and showed that in the presence 
of a background continuum, linear polarization profiles with both core and 
wing maxima are formed in isothermal as well as in chromospheric-like 
model atmospheres. 

Later, \citet{mf87,mf88} studied the linear polarization of resonance lines 
formed in isothermal slabs of finite thickness, in semi-infinite isothermal 
atmospheres and in chromospheric-like model atmosphere. She considered both 
angle-dependent and AA type II redistribution functions of 
\citet{hum62}. She confirmed the conclusion of \citet{dumontetal77}, 
namely, that the approximation of CRD is adequate to describe the 
line core polarization even when 
type II redistribution is used. Using the actual functional form of 
$R_{\rm II,AA}$, \citet{mf87,mf88} demonstrated that Kneer's approximation 
used by \citet{rs82} and \citet{gs85,gs86} cannot 
account for the phenomenon of diffusion in frequency which takes place in 
the line wings. She showed that Kneer's approximation leads to non-negligible 
errors in the wing polarization obtained from slabs of finite thickness. 
In the case of a semi-infinite atmosphere this approximation affects 
both the line core and line wing polarization, as well as the intensity itself. 
Furthermore, she showed that in the presence of a background continuum, Kneer's 
approximation predicts incorrect values for both the position and the 
magnitude of the polarization maximum in the wings. \citet{mf88} considered also
the influence of $\epsilon$ and of the line strength on the linear polarization profiles and gave a qualitative 
interpretation of the behavior of the emergent polarization profiles in terms of an Eddington-Barbier 
expression.

The collisional redistribution matrix given by \citet{dh88} was used in 
polarized line transfer computations by \citet[][in plane-parallel atmospheres]{mf92} 
and by \citet[][in spherically-symmetric atmospheres]{knn94,knn95}. 
\citet{mf92} showed that the line-wing polarization is sensitive to 
the elastic collisional rate $\Gamma_{\rm E}$ while \citet{knn94} emphasized 
that the line core polarization is sensitive to the depolarizing collisions parameter  
$D^{(2)}$. \citet{knn94,knn95} presented a detailed study about the influence of 
$\epsilon$, the damping parameter $a$, the elastic collisional rate $\Gamma_{\rm E}$ 
and $D^{(2)}$ on the linear polarization profiles, interpreting them using 
simple asymptotic expression \citep[see e.g.,][]{hubeny85a,hf80}. 

In all the above-mentioned references, Feautrier's method was used to solve the 
PRD polarized line transfer equation, except in 
\citet{knn94,knn95} who used a discrete space method. As is well-known, these type of numerical methods 
are computationally expensive. Over the last two decades fast iterative 
methods based on operator-perturbation have been developed to solve 
the PRD radiative transfer problem of resonance line polarization assuming a two-level model atom
without lower-level atomic polarization \citep[see the review by][]{knnandms09}. 
More recently, \citet{samandjtb10} have developed very efficient and 
accurate symmetric Gauss-Seidel and Successive-Overrelaxation iterative 
methods to solve the above-mentioned PRD problem. Applying this fast radiative 
transfer method, in this paper we study in detail the effects of 
various atmospheric parameters on the linear polarization profiles formed 
under PRD conditions, but using semi-empirical models of the solar atmosphere and
paying particular attention to understanding the shape of the emergent fractional linear polarization
profiles in terms of the radiation field anisotropy within the medium.
In particular, we consider two one-dimensional (1D) models of the quiet solar 
atmosphere:  one based on the VALC model of 
\citet[][which has a relatively hot lower chromosphere]{val81} and the second 
based on the MCO model of \citet[][which has a relatively cool lower chromosphere]{avrett95}. 
Although the solar atmosphere is highly inhomogeneous and 
dynamic and such 1D models can only be considered as illustrative of the 
complex atmospheric conditions, they are suitable for achieving the main aim 
of this paper, namely a basic investigation on the impact of PRD effects on 
the shape of the emergent $Q/I$ profiles and their  sensitivity to various 
atmospheric parameters. In particular, we study the sensitivity of the linear 
polarization profiles to the following relevant quantities: 
(a) the photon destruction probability per scattering 
event $\epsilon$, (b) the strength of the spectral line, parameterized as $r$ 
(see the discussion below Equation~(\ref{sk0}) for the definition of $r$), 
(c) the damping parameter $a$ of the line absorption profile, and (d) the 
elastic collisional rate $\Gamma_{\rm E}$. 

The paper is organized in the following manner\,: in Section~2 we present 
a historical background on the PRD matrix 
for non-magnetic resonance scattering in a two-level atom. The 
formulation of the problem is presented in Section~3. A detailed study of 
the sensitivity of the linear polarization profiles to various atmospheric 
parameters and to the thermal structure of the solar atmosphere is presented 
in Section~4. Concluding remarks are given in Section~5. 

\section{Redistribution Matrix for the Non-magnetic Resonance Scattering Problem}
In this section we present a brief background on the PRD 
matrix for resonance line polarization. 
We follow the same notation as in \citet{samandjtb10}. 

The two-level atom PRD problem without polarization was formulated  
by \citet{hum62}. Later, starting from the work of \citet{osc72}, more 
general PRD functions including a better 
treatment of collisions, for both resonance and subordinate lines, 
were derived by \citet{heinzel81}, \citet{hubeny82}, 
\citet{hos83a,hos83b}, \citet{hc86} and \citet{hl95}. 
Reviews on the unpolarized PRD problem and its application to astrophysics were 
presented by \citet{hubeny85b} and \citet{hf88}.
For a more recent review on the same topic, but with emphasis on numerical 
methods and radiative transfer modeling, see \citet{rh03}. 

When the polarization state of the spectral line radiation is taken into 
account the unpolarized PRD functions become $4\times 4$ redistribution 
matrices that describe how the Stokes vector is redistributed in both frequency 
and angle. In the non-magnetic case the redistribution matrix, once 
averaged over the Maxwellian distribution of the scattering atoms, 
can be expressed as a product of the angle and frequency dependent PRD 
function of \citet{hum62} times 
a frequency-independent $4 \times 4$ phase matrix. 
Clearly, there is a 
intricate coupling between the frequency, angle and polarization state of 
the radiation field. 

A $4\times 4$ phase matrix for the 
$J_l=0 \to J_u=1 \to J_l=0$ scattering transition was derived by 
\citet{chandra50} using classical electrodynamics, which is referred 
to as the Rayleigh phase matrix. Using quantum mechanics 
\citet{hamilton47} derived a more general Rayleigh phase matrix for a 
$J_l \to J_u \to J_l$ transition with arbitrary values 
for the lower and upper level angular momentum quantum numbers. For 
the particular case of a 1D atmosphere, the $4\times 4$ 
phase matrix reduces to a $2\times 2$ phase matrix, due to the azimuthal 
symmetry of the problem. 

In polarized radiative transfer, using redistribution matrices that have 
intricate couplings between frequency and angles is numerically expensive. 
Hence, to reduce the numerical work, \citet{rs82} introduced the 
so-called hybrid approximation, in which it is assumed that the $4 \times 4$ 
redistribution matrix, depending on directions and frequencies, can be 
factorized into a $4 \times 4$ matrix, depending only on directions, 
times a scalar function depending only on frequencies (the so-called 
AA redistribution function). Indeed such a factorization is valid only 
in the atomic frame. In the laboratory frame it is mainly used as a 
practical ``ansatz'' for avoiding extremely time-consuming calculations and 
is justified only by means of heuristic arguments.
Under the hybrid approximation, in a plane 
parallel atmosphere, the redistribution matrix can be written, in 
general, as a linear combination of terms of the form\footnote{Note that 
the only difference with Equations~(33) and (37) of \citet{samandjtb10} 
is that in Equations~(\ref{hybrid_approx}) and (\ref{dh_redist}) of the 
present paper we have written $\phi_x g^k_{xx^\prime}$ instead of 
$g^k_{xx^\prime}$, in order to be consistent with the definition of ${\bf R}$ 
given below.} 

\begin{equation}
{\bf R}(x,x^\prime; \mu, \mu^\prime) = (1-\epsilon)\, \phi_x\,g^{k}_{xx^\prime}\,
{\bf P}(\mu,\mu^\prime).
\label{hybrid_approx}
\end{equation}
Here $x$, $x^\prime$ are the non-dimensional frequencies of outgoing and 
incoming photons, and $\mu,\mu^\prime$ are the cosine of the polar angles
$\theta$, $\theta^\prime$ with respect to the vertical direction in 
the atmosphere. 
${\bf R}(x,x^\prime; \mu, \mu^\prime)$ gives the joint probability of absorbing 
a photon with frequency $x^\prime$ and direction $\mu^\prime$ and re-emitting 
by spontaneous de-excitation a photon with frequency $x$ and direction $\mu$. 
The photon destruction probability is given by 
$\epsilon=\Gamma_{\rm I}/(\Gamma_{\rm I}+\Gamma_{\rm R})$ with 
$\Gamma_{\rm I}$ being the inelastic de-excitation collision rate and 
$\Gamma_{\rm R}$ the radiative de-excitation rate. In 
Equation~(\ref{hybrid_approx}), $g^k_{xx^\prime}=R_{k,{\rm AA}}(x,x^\prime)/\phi_x$,
where $R_{k, {\rm AA}}$ (with $k={\rm I,\,II,\ and\ III}$) are the
AA redistribution functions of \citet{hum62} and $\phi_x$ is 
the normalized Voigt profile function. The function $R_{k, {\rm AA}}/
\phi_x$ gives the probability of absorption at frequency $x^\prime$, per 
emission at frequency $x$.  
Note that in Equation~(\ref{hybrid_approx}) we introduce for convenience the 
quantity $g^k_{xx^\prime}$ which has the advantage of formally simplifying the 
expression for the line source vector (see 
Equations~(\ref{slk0}) and (\ref{jbark0}) below). Finally, for the 
case of a 1D atmosphere, ${\bf P}(\mu,\mu^\prime)$ denotes the 
$2\times 2$ phase matrix. 

The redistribution matrix given in Equation~(\ref{hybrid_approx}) does not 
take into account the elastic collisions, quantified by the parameters 
$\Gamma_{\rm E}$ and $D^{(2)}$. However, the elastic collisional rate 
$\Gamma_{\rm E}$ 
can be included into the hybrid approximation by replacing 
$g^{k}_{xx^\prime}$ by a linear combination of type II and type III 
PRD functions, weighted by the factor 
$\gamma=(\Gamma_{\rm R}+\Gamma_{\rm I})/(\Gamma_{\rm R}+\Gamma_{\rm I}+\Gamma_{\rm E})$, namely
\begin{equation}
{\bf R}(x,x^\prime; \mu, \mu^\prime) = (1-\epsilon)\, \phi_x\,
\left[\gamma g^{\rm II}_{xx^\prime} + 
(1-\gamma)g^{\rm III}_{xx^\prime}\right]
{\bf P}(\mu,\mu^\prime).
\label{r23_mix}
\end{equation}
We remark that the $(1,1)$ element of ${\bf R}$ given in 
Equations~(\ref{hybrid_approx}) and (\ref{r23_mix}) is normalized to 
$(1-\epsilon)$, because it does not take into account the 
depolarizing collisional rate $D^{(2)}$. 
We refer the reader to \citet{osc72} for a detailed discussion on the 
normalization of the redistribution matrix.
To qualitatively model the solar Ca\,{\sc ii} K line \citet{gs85} 
used the redistribution matrix given by Equation~(\ref{r23_mix}), but with a 
further simplification which 
consisted in replacing $R_{\rm II,AA}$ by Kneer's (1975) approximation and 
$R_{\rm III,AA}$ by the CRD expression. Also, he neglected $\Gamma_{\rm I}$ in the 
expression for $\gamma$ (see above), and assumed that the matrix 
${\bf P}(\mu, \mu^\prime)$ which multiplies $R_{\rm III,AA}$ 
(see Equation~(\ref{r23_mix})) is the isotropic scattering phase matrix.
Moreover, \citet{gs85} did not take into account the 
depolarizing collisional rate $D^{(2)}$ in his modeling. 

The redistribution matrix given by \citet{dh88} for the unmagnetized 
case \citep[see also][]{vb97b} takes into account the effect of depolarizing collisions 
through the $D^{(2)}$ parameter. It is worth noting that this collisional redistribution 
matrix is very general, namely, it depends  
on the angle-dependent redistribution functions of \citet{hum62}. However, 
for computational simplicity, following \citet{rs82}, some authors have used 
the AA version of this general collisional redistribution matrix 
\citep[e.g.,][]{mf92,knn94}. Following \citet{vb97b} the 
AA version of the collisional redistribution matrix can be written as 
\citep[see also Equation~(37) of][]{samandjtb10}
\begin{eqnarray}
&{\bf R}(x,x^\prime; \mu,\mu^\prime)=  
\sum_{K=0,2} W_{K}(J_l,J_u) \nonumber \\ 
&\times \,\phi_x\,\left\{\alpha\, g^{\rm II}_{xx^\prime} + 
[\beta^{(K)}-\alpha]\, g^{\rm III}_{xx^\prime}\right\}
{\bf P}^{K}_{\rm R}(\mu,\mu^\prime),
\label{dh_redist}
\end{eqnarray}
where the branching ratios $\alpha$ and $\beta^{(K)}$ are given by
\begin{equation}
\label{branch_alpha}
\alpha={\Gamma_{\rm R}\over \Gamma_{\rm R}+\Gamma_{\rm I}+\Gamma_{\rm E}},
\end{equation}
\begin{equation}
\beta^{(K)}={\Gamma_{\rm R}\over \Gamma_{\rm R}+\Gamma_{\rm I}+D^{(K)}}.
\label{branch_beta}
\end{equation}
Note that $D^{(0)}=0$, and also that the factor $(1-\epsilon)$ is contained
in the branching ratios. The coefficient $W_{0}(J_l,J_u)=1$, and 
$W_{2}(J_l,J_u)$ characterizes the maximum linear polarization
that can be produced in the line. In the case of a normal Zeeman triplet
($J_l=0, J_u=1$), $W_{2}(J_l,J_u)=1$. Furthermore, 
${\bf P}^{K}_{\rm R}(\mu,\mu^\prime)$
are the Rayleigh phase matrix multipolar components 
\citep[][see also Equation~(35) of Sampoorna \& Trujillo Bueno 2010]{lan84}. 

The CRD limit is obtained from Equation~(\ref{dh_redist}) by taking  
$\Gamma_{\rm E} \gg \Gamma_{\rm R}$ (in other words $\alpha \ll 1$)
and $g^{\rm III}_{xx^\prime}=\phi_{x^\prime}$ 
\citep[compare the resulting expression with Equation~(10.54) of][with the magnetic field strength set to zero]{ll04}. 
Notice that CRD means that there is no frequency correlation between 
the incoming and outgoing photons. In other words, the incident radiation is 
completely redistributed in frequency, so that there is complete non-coherence 
between the incident and scattered photon frequencies. This CRD limit is reached
when the atomic system is illuminated by a spectrally flat radiation field \citep[e.g.,][]{ll04}. 
Another good approximation 
to this CRD limit occurs when the atoms are so strongly perturbed by elastic collisions 
during the scattering process that the excited electrons are randomly 
redistributed over the substates of the upper level \citep[see][]{mih78}. 
However, in the context of polarization, adopting the CRD approximation 
appears to lead to the following contradiction\,: CRD is obtained in the limit of 
very high collisional rates with respect to the radiative rate, but then 
the spectral line would be completely depolarized by 
depolarizing collisions. This apparent contradiction can be clarified by 
recalling the explicit physical meaning of the branching ratios appearing in 
Equation~(\ref{dh_redist}). First of all, for the problem at hand, elastic 
collisions produce two different effects represented by the collisional 
rates $\Gamma_{\rm E}$ and $D^{(2)}$. The collisional rate $\Gamma_{\rm E}$ 
is responsible for line broadening and for destruction of correlations between incoming 
and outgoing photon frequencies. The rate $D^{(2)}$ is responsible for the 
destruction of the upper-level atomic alignment and, thereby, causes depolarization. 
The branching ratio $\alpha$ gives the probability that a radiative decay from the  
excited state occurs before any type of collision (elastic or inelastic). 
Therefore, $\alpha$ gives the fraction of the scattering processes that are 
coherent in the atomic rest frame ($R_{\rm II}$). The branching ratio 
$\beta^{(K)}$ gives the probability that radiative decay of the excited state 
occurs without destruction of the $2K$-multipole moment. Therefore, the total 
branching ratio $\left[\beta^{(K)}-\alpha\right]$ gives the probability that 
radiative de-excitation occurs after a elastic collision ($\Gamma_{\rm E}$) 
which redistributes the radiation in frequency without destroying the 
$2K$-multipole moment. Clearly $\left[\beta^{(K)}-\alpha\right]$ gives the 
fraction of the scattering processes that are completely non-coherent in the 
atomic rest frame ($R_{\rm III}$). Therefore, to get the CRD limit we only 
assume that elastic collisions that completely redistribute the radiation 
in frequency are sufficiently strong 
($\Gamma_{\rm E}\gg \Gamma_{\rm R}$), so that the branching ratio 
$\alpha \to 0$. In other words, $D^{(2)}$ and $\Gamma_{\rm E}$ are assumed 
to be independent, even though their physical origin are the elastic 
collisions. In this case Equation~(\ref{dh_redist}) reduces to 
\begin{eqnarray}
&{\bf R}_{\rm CRD}(x,x^\prime; \mu,\mu^\prime)=  
\sum_{K=0,2} W_{K}(J_l,J_u) \nonumber \\ 
&\times \,\beta^{(K)}\phi_x\, \phi_{x^\prime}
{\bf P}^{K}_{\rm R}(\mu,\mu^\prime),
\label{crd_redist}
\end{eqnarray}
where we have further replaced $g^{\rm III}_{xx^\prime}$ by $\phi_{x^\prime}$. 
Clearly the branching ratio $\beta^{(K)}$ allows only that fraction of the 
scattering processes which do not lead to destruction of the $2K$-multipole 
moment. 

We compute the $R_{\rm II,AA}$ and $R_{\rm III,AA}$ 
functions by numerically integrating the corresponding angle-dependent 
redistribution functions \citep[see Equations~(59) and (61) of][]{vb97b} over 
all the scattering angles \citep[see Equations~(103) and (104) of][]{vb97b}. 
For the numerical integration over all the scattering angles we have used a 
15 point Gauss-Legendre quadrature. In 
this paper we assume that the redistribution functions $R_{\rm II,AA}$ and 
$R_{\rm III,AA}$ and the Voigt profile function $\phi_x$ are constant 
throughout the atmosphere. Since these functions depend on the damping 
parameter $a$ and on the reduced frequency $x$, 
which is defined as $x = (\nu_0-\nu)/\Delta \nu_{\rm D}$, with $\nu_0$ the line 
center frequency and $\Delta \nu_{\rm D}$ the Doppler width, the above 
assumption implies that $a$ and $\Delta \nu_{\rm D}$ are also constant 
within the entire atmosphere. 

\section{Formulation of the Problem}
In this section we discuss all the physical ingredients required for 
our investigation (presented in Section~4) about PRD scattering polarization profiles. 
Section~3.1 describes the basic equations of the problem. The model atmospheres 
used for our study in Section~4 are discussed in Section~3.2. 
The definition of the anisotropy factor for PRD is considered in Section~3.3, 
where we also discuss several known facts about the anisotropy and its 
relation to the source function gradient. In Section~3.4 we present 
the Eddington-Barbier relation for PRD problems, and show the direct dependence 
of the emergent polarization on the anisotropy within the model atmosphere under consideration. 

\subsection{The Basic Equations}
We consider the standard two-level atom resonance line polarization 
problem in a one-dimensional, plane-parallel, static stellar atmosphere. 
Furthermore, we assume an unpolarized background continuum 
with no continuum scattering. The basic equations for the 
above-mentioned problem have been presented 
in detail in \citet[][see their Section~5]{samandjtb10}. Here we follow the 
same formulation, but for the sake of clarity we recall a few 
important basic equations from that paper. 

As shown by \citet{chandra50}, an azimuthally symmetric polarized radiation 
field can be described by the two Stokes parameters $I_{x\mu}$ and $Q_{x\mu}$. 
The Stokes $I_{x\mu}$ parameter is the specific intensity, while $Q_{x\mu}$ 
quantifies the linear polarization (i.e., the difference between the intensity 
components parallel and perpendicular to a given reference direction in the 
plane perpendicular to the direction of the ray under consideration). In this 
paper, the positive $Q_{x\mu}$ direction is defined in the plane containing 
the direction of the ray and the vertical Z-axis. When polarization is taken 
into account, the Stokes source vector depends not only on the frequency $x$ 
but also on $\mu=\cos \theta$, with $\theta$ the angle between the ray and the 
vertical Z-axis. It is possible to transform 
from the Stokes basis to an irreducible basis 
where the source vector components depend only on the frequency $x$. Such a 
transformation is referred to as the ``decomposition'' of the Stokes vector. 
\citet{hf07} has given a simple way of achieving this goal using the 
irreducible tensors for polarimetry introduced by \citet{lan84}. This 
decomposition technique was used in \citet{samandjtb10}, 
and we adopt it here also. 

In the reduced basis the irreducible intensity $(I_{x\mu})^K_0$ with 
$K=0\ {\rm and}\ 2$ satisfies the following transfer equation
\begin{equation}
{{\rm d}  \over {\rm d} \tau_{x\mu}}({I}_{x\mu})^K_0(\tau_{x\mu}) =
({I}_{x\mu})^K_0(\tau_{x\mu}) - ({S}_{x})^K_0(\tau_{x\mu}),
\label{prte-ik0}
\end{equation}
where the irreducible source vector components are given by 
\begin{equation}
({S}_{x})^K_0 = {\phi_x ({S}_{lx})^K_0 + r B{U}^K_0\over \phi_x+r} .
\label{sk0}
\end{equation}
Here $\tau_{x\mu}$ is the total optical depth defined by 
${\rm d}\tau_{x\mu}= -(\chi_l\phi_x+\chi_c){\rm d}z/\mu$, with $z$ the 
distance along the normal to the atmosphere, $\chi_{l}$ and $\chi_c$ are 
the line and continuum opacities, $r = \chi_c/\chi_l$,  
$B$ is the Planck function, and $U^K_0$ is equal to 
unity for $K=0$ and zero for $K=2$. The irreducible line source vector 
components are given by 
\begin{equation}
({S}_{lx})^K_0=\epsilon B {U}^K_0 
+  W_K(J_l,J_u) (\bar J_x)^K_0,
\label{slk0}
\end{equation}
where
\begin{equation}
(\bar J_x)^K_0 = \int_{-\infty}^{+\infty} {\rm d}x^\prime 
\left\{\alpha g^{\rm II}_{xx^\prime}
+\left[\beta^{(K)}-\alpha\right] g^{\rm III}_{xx^\prime}\right\} 
(J_{x^\prime})^K_0. 
\label{jbark0}
\end{equation}
The angle integrated irreducible tensors of the radiation field are 
respectively given by
\begin{equation}
(J_x)^0_0 = {1 \over 2} \int_{-1}^{+1} {\rm d}\mu\, I_{x\mu},
\label{j00}
\end{equation}
\begin{equation}
(J_x)^2_0 = {1 \over 4\sqrt{2}} \int_{-1}^{+1} {\rm d}\mu\, 
\left[(3\mu^2-1)I_{x\mu} + 3(\mu^2-1)Q_{x\mu}\right].
\label{j20}
\end{equation}
The irreducible components of the radiation field and of the source 
function are connected to the Stokes parameters and to the source 
vector through a simple formula
\citep[see Appendix~B of][]{hf07}. Note that in the solar atmosphere 
the dominant contribution 
to $(J_x)^2_0$ is given by the first term in the square bracket of 
Equation~(\ref{j20}).

\subsection{The Model Atmospheres}
\begin{figure}
\plotone{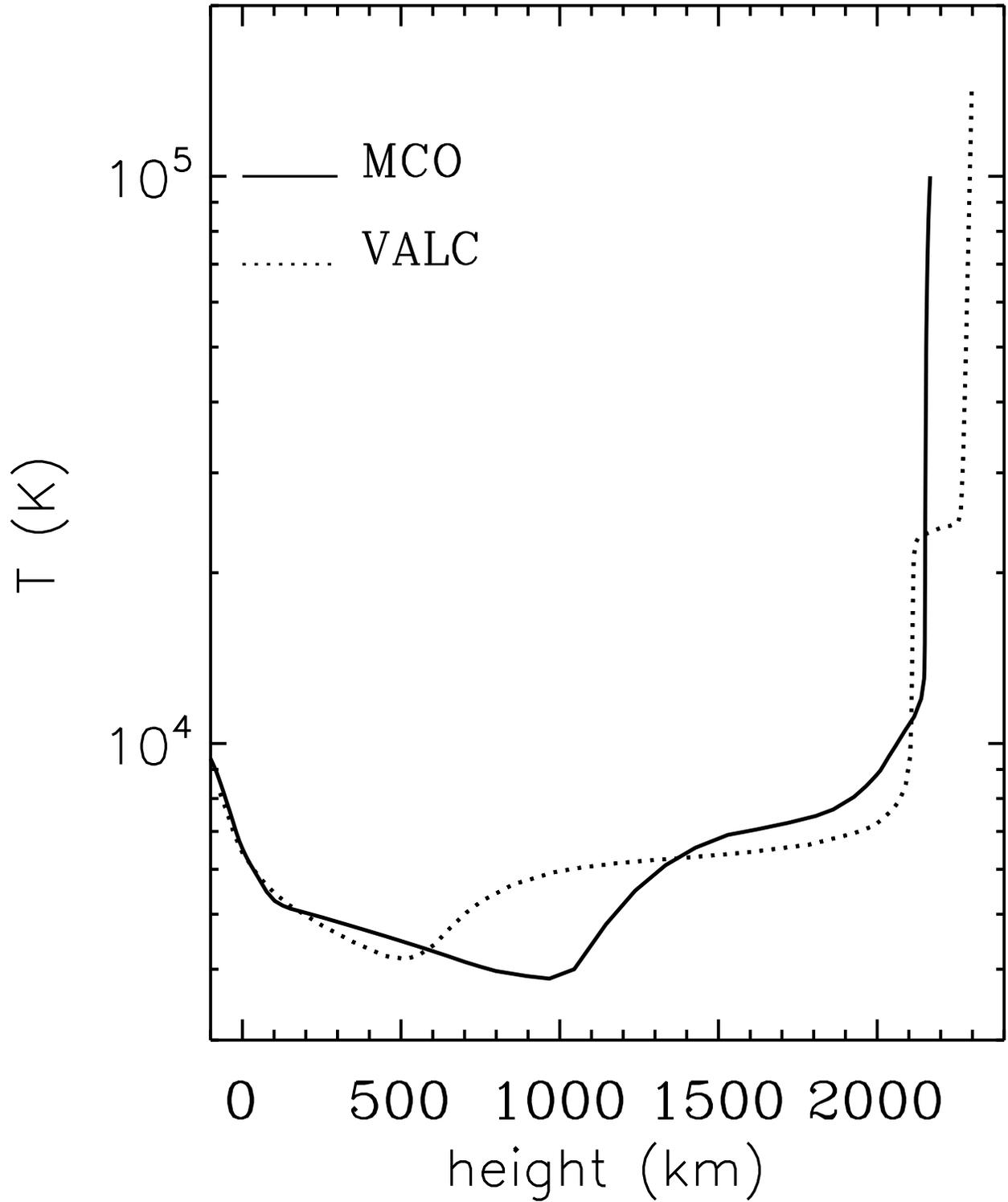}
\caption{Temperature stratification of the VALC (dotted line) and MCO 
(solid line) solar model atmospheres. Note that between 600 and 1300 km 
(hereafter, the ``lower chromosphere") the MCO model is cooler than VALC.
}
\label{valcandmco}
\end{figure}

In this section we describe the two solar model atmospheres that we have 
chosen for our detailed study of the linear polarization profiles presented 
in Section~4. As already mentioned in Section~1, we consider the 
VALC and the MCO models for this purpose. Figure~\ref{valcandmco} shows 
the variation of the temperature as a function of height for the VALC (dotted 
line) and MCO (solid line) models. We note that both models 
include the transition region as well as the corona. Most 
of the results presented in this paper correspond to spectral lines 
which become optically thin for heights $z>2100$ km. However, we  
also present some results for a very strong spectral line whose line center originates in the 
transition region of the VALC model. 

For a hypothetical line at $\lambda = 5000$ \AA\ we calculate the Planck 
function $B$ (in erg\,s$^{-1}$\,cm$^{-2}$\,sr$^{-1}$\,cm$^{-1}$) 
as\footnote{Rigorously speaking, the Planck function expression should be that corresponding to the 
Wien limit because we are neglecting stimulated emission. However, for visible and UV lines both expressions give similar values.}  
\begin{equation}
B = {2 h c^2 \over \lambda^5} \, { 1 \over {\rm e}^{hc /
(\lambda k_{\rm B} T)} -1},
\label{planck}
\end{equation}
where $c$ is the speed of light, 
$h$ is the Planck constant, and $k_{\rm B}$
is the Boltzmann constant. As the tabulated height 
grid of the VALC and MCO models is crude, we actually first interpolate the 
tabulated temperatures on a much finer height grid. We use cubic spline 
interpolation for this purpose. 
Our height grid has a spacing of $\Delta z=5$\,km. Note that achieving the accuracy corresponding to  
such a fine spatial grid requires finding the self-consistent solution via the application of a 
highly-convergent iterative scheme, such as that on which the PRD computer program 
used here is based on \citep[see][]{samandjtb10}.

\begin{figure}
\plotone{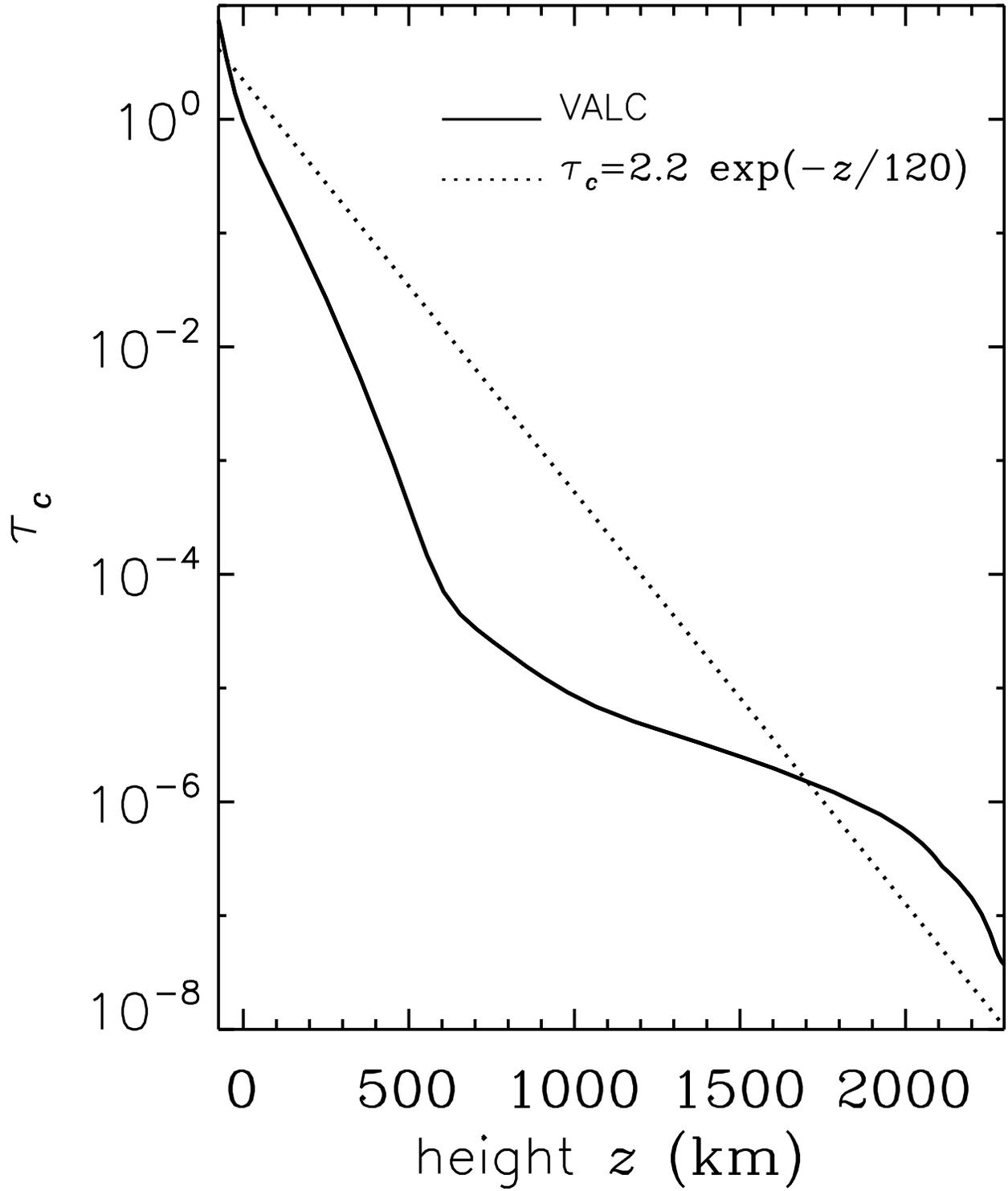}
\caption{The continuum optical depth, $\tau_c$ versus height in the VALC 
model atmosphere (solid line) and for a 
exponentially stratified atmosphere (dotted line). 
In this paper we have used 
$\tau_c=2.2\,{\exp}(-z/H)$, with $H=120$\,km 
and $z$ in km. 
}
\label{valctaucvsexptauc}
\end{figure}

We assume that both solar model atmospheres are exponentially stratified, 
so that the continuum optical depth is described by the 
law $\tau_c=2.2\,\exp(-z/H)$ with $z$ (in km) the vertical 
height in the atmosphere, and $H$ the scale height (which we have chosen 
equal to 120\,km). Moreover, we assume that the line strength parameter $r$, 
the photon destruction probability per scattering $\epsilon$, and the damping 
parameter $a$ of the line absorption profile are constant with height. 
Figure~\ref{valctaucvsexptauc} 
shows a plot of $\tau_c$ versus height for the VALC model. For $\tau_c$ 
we use the tabulated value at $\lambda=5000$\,\AA\ from \citet{val81}. 
Overplotted is our continuum optical depth versus height. 
Clearly, the $\tau_c$ of the VALC model does not show a single 
scale height. In fact, our attempts to fit the $\tau_c$ of VALC required the use of a weighted combination of 
two exponentials with two different scale heights. We find that for 
heights lower than 600\,km one has to use a scale height of 60\,km, while for 
larger heights one needs a scale height of 250\,km, in order to reasonably fit the 
$\tau_c$ of VALC. However, to keep the problem as simple as possible, we choose 
a single scale height of 120\,km for the entire atmosphere and $\tau^c_0=2.2$ 
that nearly coincides with the $\tau_c$ value of the VALC model at the lower 
boundary of the atmosphere (see dotted line in Figure~\ref{valctaucvsexptauc}). 
The same choice applies to the MCO model.
This strategy appears to be reasonable for facilitating a systematic 
investigation on the impact of PRD effects on the shape of the emergent $Q/I$ profiles.

\subsection{The Anisotropy}
The $(J_x)^2_0$ tensor, which is dominated by the contribution from 
the Stokes $I_{x\mu}$ parameter, characterizes the `degree of anisotropy' 
of the radiation field \citep[][see also Trujillo Bueno 2001]{ll04}. 
We define a `monochromatic anisotropy factor' as 
\begin{equation}
A=(J_x)^2_0/(J_x)^0_0.
\label{monochromatic_aniso}
\end{equation}
For clarity of notation, we denote the 
monochromatic anisotropy $A$ for CRD and PRD by $A_{\rm CRD}$ and 
$A_{\rm PRD}$, respectively. 
We also define a `frequency integrated or mean anisotropy factor' for PRD as 
\begin{equation}
\bar A_{\rm PRD}={(\bar J_x)^2_0 / (\bar J_x)^0_0},
\label{anisotropy_factor}
\end{equation}
where $(\bar J_x)^0_0$ and $(\bar J_x)^2_0$ are defined as in 
Equation~(\ref{jbark0}). 
In the case of CRD the frequency integrated anisotropy factor is 
defined by 
\begin{equation}
\bar A_{\rm CRD}={(\bar J^2_0)_{\rm CRD} / (\bar J^0_0)_{\rm CRD}},
\label{anisotropy_factor_crd}
\end{equation}
where 
\begin{equation}
(\bar J^K_0)_{\rm CRD} = \int_{-\infty}^{+\infty} \beta^{(K)} 
\phi_x (J_x)^K_0 {\rm d}x.
\label{jbark0_crd}
\end{equation}
Clearly, the mean anisotropy $\bar A_{\rm CRD}$ does not depend explicitly on 
frequency. However, in the solar atmosphere $\bar A_{\rm CRD}$ 
changes with height. It then results that, even in CRD, since different 
points of the profile form at different heights, we have an ``apparent'' 
frequency dependence of $\bar A_{\rm CRD}$ across the profile. On the 
contrary, $\bar A_{\rm PRD}$ depends explicitly on frequency and this effect 
adds to the previous one. 

\begin{figure*}
\plotone{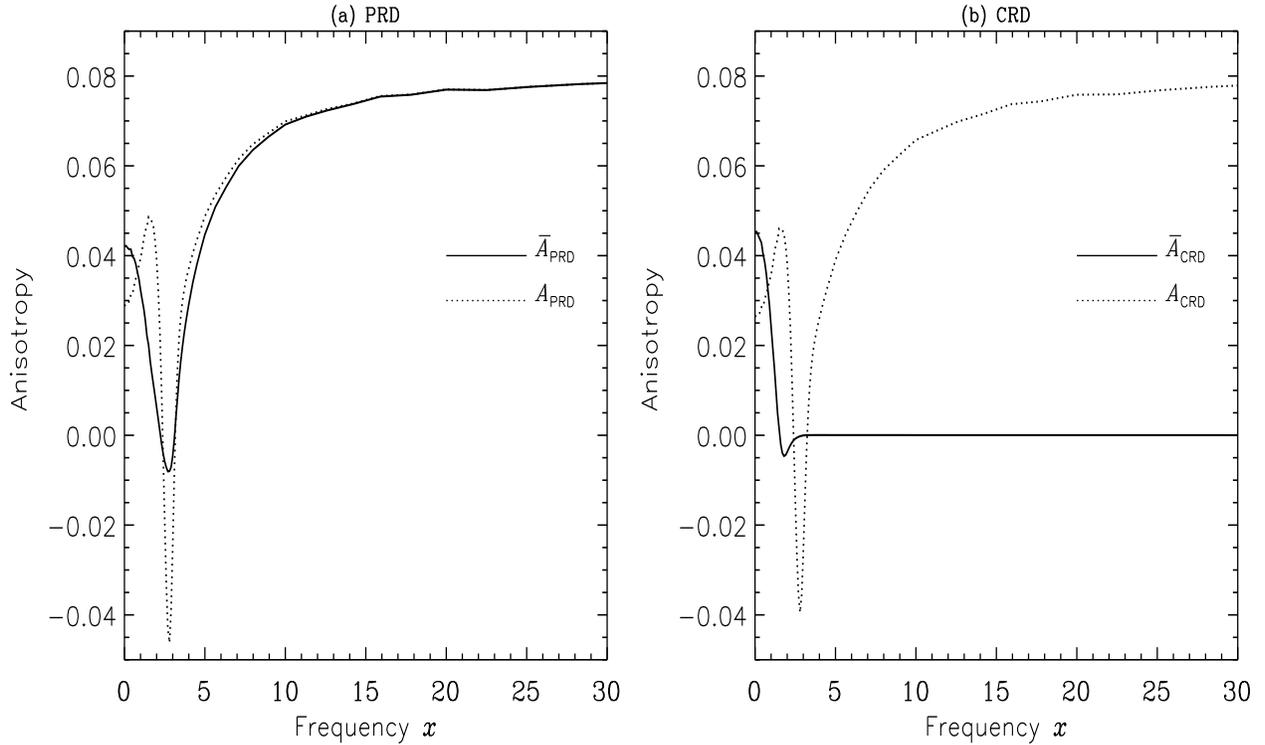}
\caption{The anisotropy versus the reduced frequency $x$ at the height in 
the VALC model atmosphere where $\tau_{x\mu}=1$ for a LOS with $\mu=0.11$. 
Panels (a) and (b) correspond to PRD and CRD respectively. 
The other model parameters are $\epsilon=10^{-4}$, $r=10^{-5}$, $a=10^{-3}$ 
and $\Gamma_{\rm E}/\Gamma_{\rm R}=D^{(2)}/\Gamma_{\rm R}=0$. In both 
panels the solid line is $\bar A$ and the dotted line is $A$. Notice that 
$A_{\rm CRD}$ and $A_{\rm PRD}$ are similar, while $\bar A_{\rm CRD}$ and 
$\bar A_{\rm PRD}$ differ greatly, particularly in the wings. 
}
\label{abarvsa_crvspr}
\end{figure*}

In Figure~\ref{abarvsa_crvspr} we illustrate the important difference between 
the monochromatic and mean anisotropy factors for both, the PRD and CRD cases. 
To compute these anisotropy factors we 
have used the VALC model atmosphere (see Section~3.2) and $\epsilon=10^{-4}$, 
$r=10^{-5}$, $a=10^{-3}$ and $\Gamma_{\rm E}/\Gamma_{\rm R}=D^{(2)}/
\Gamma_{\rm R}=0$. In Figure~\ref{abarvsa_crvspr} we plot $A$ and $\bar A$ 
versus the reduced frequency at the height where for a line-of-sight (LOS) 
with $\mu=0.11$ the condition $\tau_{x\mu}=1$ is satisfied. Clearly, the 
monochromatic anisotropy factors $A_{\rm CRD}$ and $A_{\rm PRD}$ are similar. 
The differences are basically due to the slight differences that are found 
in Stokes $I_{x\mu}$ for CRD and PRD (compare the dotted lines in the 
intensity panels of Figures~\ref{valc_prd_r} and \ref{valc_crd_r}). 
It is interesting to note that $A_{\rm CRD}(\tau_{x\mu}=1)$ does not 
tend to zero in the wings. This is because the condition $\tau_{x\mu}=1$ 
for a LOS with $\mu=0.11$ is satisfied at the height of 360\,km for all 
frequencies $x\ge 20$. At this height the anisotropy $A_{\rm CRD}$ for 
$x\ge 20$ is determined by the Planck function gradient at 
that height. The mean anisotropy factors $\bar A_{\rm PRD}$ and 
$\bar A_{\rm CRD}$ differ greatly, particularly in 
the wings. This can be understood as follows. In the case of CRD we  
integrate $(J_x)^2_0$ and $(J_x)^0_0$ over all $x$ after multiplying them 
by the absorption profile function $\phi_x$ (see Equation~(\ref{jbark0_crd})), 
which goes to zero in the wings. Thus, at any depth $\tau_{x\mu}$ the 
dominant contribution to $(\bar J^K_0)_{\rm CRD}$ and, thereby to 
$\bar A_{\rm CRD}$ comes from the line core region. 
Furthermore, the monochromatic anisotropy $A_{\rm CRD}$ decreases with depth, 
particularly in the line core. As already noted above, the frequency 
dependence of $\bar A_{\rm CRD}$ is due to the 
fact that different parts of the line profile are formed at different heights. 
Combining these facts, it is easy to see that $\bar A_{\rm CRD}$ at 
$\tau_{x\mu}=1$, decreases as the frequency increases. In particular, far in 
the wings of strong lines like the one we are considering in 
Figure~\ref{abarvsa_crvspr} (i.e., with $r=10^{-5}$) $\bar A_{\rm CRD}$ 
is zero (even if $A_{\rm CRD}$ is substantial), 
because we are looking deep in the atmosphere where $A_{\rm CRD}$ in the line 
core is zero (which dominantly contributes to $\bar A_{\rm CRD}$). 
In the case of PRD, we integrate $(J_x)^2_0$ and $(J_x)^0_0$ over all 
$x$ after multiplying them by the type II redistribution function 
(see Equation~(\ref{jbark0})), which behaves like CRD in the core and 
like coherent scattering 
in the wings. Thus, unlike CRD, at any depth $\tau_{x\mu}$ the mean anisotropy 
$\bar A_{\rm PRD}$ depends explicitly on $x$. Furthermore, in the line core it 
looks somewhat similar to $\bar A_{\rm CRD}$. But in the wings 
$\bar A_{\rm PRD}$ coincides with $A_{\rm PRD}$, due to the coherent behavior 
of $g^{\rm II}_{xx^\prime}$. 

The importance of the anisotropy factor, which is the fundamental quantity 
that determines the shape of the emergent polarization, was realized by 
\citet{rs82} who pointed out 
that the polarization is a mapping of the depth dependence of the anisotropy 
of the radiation field within the atmosphere. However, they 
neither quantified nor presented a detailed study of the anisotropy factor. 
Some information can be found in the PhD thesis of \citet{gs86}, who 
presented a study of the anisotropy for the CRD case 
(see his Chapter~4). He also illustrated the relation of the anisotropy 
factor to the source function gradient by considering, in a finite slab 
atmosphere, a simple piece-wise linear source function. His 
quantities $\alpha$ and $\bar \alpha$ are quite 
similar to our quantities $A_{\rm CRD}$ and $\bar A_{\rm CRD}$. However, 
in the case of PRD, \citet{gs86} did not define a mean anisotropy factor 
$\bar A_{\rm PRD}$ similar to that defined in our 
Equations~(\ref{anisotropy_factor}) and (\ref{jbark0}) as he used 
Kneer's approximation for $R_{\rm II,AA}$. Nevertheless, 
he correctly noted that for line core frequencies the anisotropy of the 
radiation field is controlled by $\bar A_{\rm CRD}$ 
and for the line wings it is controlled by $A_{\rm PRD}$. 
In this paper we show 
that a better way of defining the anisotropy for the PRD problem is 
through Equation~(\ref{anisotropy_factor}) along with 
Equations~(\ref{jbark0})--(\ref{j20}). 

A Milne-Eddington atmosphere is a very suitable model to clearly demonstrate 
the strong dependence of the anisotropy factor on the gradient of the source 
function for the Stokes $I_{x\mu}$ parameter \citep[see Section~3 of][]{jtb01}. The mean anisotropy $\bar A_{\rm CRD}$ is bounded by 
$-1/2 \le \sqrt{2} \bar A_{\rm CRD} \le 1$, the lower and upper bounds being 
for a purely horizontal radiation field without azimuthal dependence and 
for a purely vertical illumination, respectively. 
We summarize the following facts about the anisotropy in a stellar atmosphere\,:

\noindent
(1) The anisotropy is essentially negative for an atmosphere with no or very 
small gradient in the source function. The negative values of the anisotropy 
means that the radiation field is predominantly limb-brightened (i.e., 
predominantly horizontal). \\
(2) The larger the source function gradient the larger the anisotropy. 
Positive anisotropy values imply that the radiation 
field at the spatial point under consideration 
is predominantly limb-darkened (i.e., predominantly vertical). 

In summary, in a stellar atmosphere the outgoing radiation is 
predominantly vertical (which makes a positive contribution to 
$(\bar J_x)^2_0$), while the incoming radiation is predominantly 
horizontal (which makes a negative contribution to $(\bar J_x)^2_0$). More 
precisely, ``vertical'' rays (i.e., with $|\mu|>1/{\sqrt{3}}$) make positive 
contributions to $\bar A$, while ``horizontal'' rays (i.e., with 
$|\mu|<1/{\sqrt{3}}$) make negative contributions to $\bar A$. As seen in 
Figure 4 of \citet{jtb01}, the larger the gradient of the source function the 
greater the anisotropy factor of the pumping radiation field, and the larger 
the amount of atomic level polarization. A more detailed study of the 
anisotropy factor and its relation to the source function gradient can be 
found in \citet{ll04}. 

More recently, \citet{reneetal05} have presented a study of the monochromatic 
anisotropy $A_{\rm PRD}$ for the solar Ca\,{\sc i} 4227\,\AA\ and 
Na\,{\sc i} 5890\,\AA\ lines taking into account the AA collisional 
redistribution matrix \citep[see, e.g.,][]{jos94}. They also 
illustrate the relation of $A_{\rm PRD}$ to the source function 
gradient. However, as they consider only $A_{\rm PRD}$ the interpretation of 
the emergent linear polarization in terms of $A_{\rm PRD}$ is not 
straightforward. In this paper we show that a better quantity that can be 
directly related to the emergent linear polarization is $\bar A_{\rm PRD}$. 

\subsection{The Eddington-Barbier Relation}
In the case of CRD, the emergent 
polarization directly depends on the anisotropy of the radiation field 
\citep[see Equation~(13) of][see also Section~1 of Trujillo Bueno 2003]{jtb99}. 
Such a relation can be considered as the 
generalization of the Eddington-Barbier relation for scattering polarization. 
Thus, the emergent polarization is governed by the anisotropy of the radiation 
field, which in turn is strongly related to the gradient of the source function 
for Stokes $I_{x\mu}$ and, therefore, to the $\mu$ dependence of $I_{x\mu}$. 
Note that the Eddington-Barbier relation given by \citet{jtb99,jtb03} for 
the CRD problem is very general, as it takes into account the effects of 
lower-level polarization. 

An Eddington-Barbier relation for the PRD case was given by 
\citet{mf87,mf88} who used it for a qualitative analysis of the 
linear polarization profiles formed in an isothermal model atmosphere. 
According to the Eddington-Barbier relation the emergent intensity and 
the emergent polarization from a semi-infinite atmosphere can be written as 
\citep[see also Equation~(41) of][]{hfetal09}
\begin{eqnarray}
&I_{x\mu}(\tau_{x\mu}=0)\simeq(S_{x})^0_0(\tau_{x\mu}=1), \nonumber \\
&\!\!\!\!\!\!\!Q_{x\mu}(\tau_{x\mu}=0)\simeq{3\over 2\sqrt{2}}(\mu^2-1)(S_{x})^2_0(\tau_{x\mu}=1). 
\label{eb_prd}
\end{eqnarray}
Restricting to the pure line case, and neglecting the thermal term 
$\epsilon B$, it is easy to show from 
Equations~(\ref{slk0})--(\ref{j20}) and (\ref{anisotropy_factor}), that 
\begin{equation}
\left({Q_{x\mu} \over I_{x\mu}}\right)(\tau_{x\mu}=0) \simeq {3\over 2\sqrt{2}}
(\mu^2-1)\, W_2\,\bar A_{\rm PRD}(\tau_{x\mu}=1).
\label{qoveri_eb}
\end{equation}
This approximate formula shows clearly that the emergent fractional linear polarization 
depends on the mean anisotropy $\bar A_{\rm PRD}$ at $\tau_{x\mu}=1$ along 
the LOS. In this paper, we therefore show $\bar A_{\rm PRD}$ versus the 
reduced frequency at the height in the model atmosphere where $\tau_{x\mu}=1$ 
for a LOS with $\mu=0.11$. 
Furthermore, as our reference direction for positive Stokes $Q$ is defined perpendicular to the limb, we plot 
$-Q/I$ whose positive values indicate polarization parallel to the limb. 

\section{The Emergent Linear Polarization Profiles}
In this section we present a detailed study of the PRD scattering polarization 
profiles computed in the VALC and MCO model atmospheres (see Section~3.2 for 
details on the assumptions made). For these two atmospheric models 
we present our results by giving, for each choice of atmospheric parameters, 
a set of figures with four panels showing\,: 
(a) the emergent intensity at $\mu=0.11$ (normalized to the continuum 
intensity) as a function of the reduced frequency; (b) the emergent fractional 
linear polarization at $\mu=0.11$ as a function of the reduced frequency; (c) 
the mean anisotropy factor, $\bar A$, evaluated at the height where the 
monochromatic optical depth along a LOS with $\mu=0.11$ is equal to unity, 
as a function of 
reduced frequency; (d) the line source function at line center $(x=0)$ for the 
intensity propagating along a LOS with $\mu=0.11$ (denoted as $S_I$) as a 
function of height. For the sake of comparison, the corresponding CRD linear 
polarization profiles are shown for certain cases. 

\subsection{Results of calculations in the VALC Model}
In this section we study the influence of the line strength parameter $r$, the 
photon destruction probability per scattering $\epsilon$, the damping 
parameter $a$, the elastic collisional rate $\Gamma_{\rm E}$ and the 
depolarizing rate $D^{(2)}$ on 
the PRD linear polarization profiles. We use $\lambda=5000$\,\AA,\ 
$\epsilon=10^{-4}$, $r=10^{-5}$, $a=10^{-3}$, and $\Gamma_{\rm E}/
\Gamma_{\rm R}=D^{(2)}/\Gamma_{\rm R}=0$ as the nominal case, around 
which we vary the various parameters mentioned above. Such a study should help 
us to understand the results of future, more realistic computations where all 
the above-mentioned 
parameters are depth dependent. Unless stated otherwise we set $W_2=1$, 
which represents the case of a line transition with $J_l=0$ and $J_u=1$. For 
the sake of clarity we divide this section about the results of our radiative transfer 
calculations in the VALC model in various subsections dealing with 
the sensitivity of the solution to $r$ (Section~4.1.1),  $\epsilon$ (Section~4.1.2), 
$a$ (Section~4.1.3), $\Gamma_{\rm E}/\Gamma_{\rm R}$ and 
$D^{(2)}/\Gamma_{\rm R}$ (Section~4.1.4). Some additional studies are presented in 
Section~4.1.5 concerning lines with wavelengths other than 
$\lambda=5000$\,\AA,\ lines formed in the transition 
region, and lines with other $W_2$ values. 

\subsubsection{Effect of the Line Strength Parameter $r$}
Figure~\ref{valc_prd_r} shows the response of $I(\tau_{x\mu}=0)$, 
$-(Q/I)(\tau_{x\mu}=0)$, $\bar A_{\rm PRD}(\tau_{x\mu}=1)$, and 
$S_{\rm I}(x=0)$ for a LOS with $\mu=0.11$ to variations in $r$. For comparison the 
corresponding CRD case is shown in Figure~\ref{valc_crd_r}. For notational 
simplicity we drop the subscripts $x\mu$ on $I$ and $Q$. 
Table~\ref{formation_height_valc} 
gives the height in the VALC model atmosphere at which lines of different 
strengths have $\tau_{0\mu}=1$, both for $\mu=0.11$ and $\mu=1$. 

We first discuss the PRD line profiles presented in Figure~\ref{valc_prd_r} 
and then compare them with the CRD profiles of 
Figure~\ref{valc_crd_r}. Table~\ref{formation_height_valc} indicates that as 
$r$ decreases we progressively sample lines that are formed higher in the 
atmosphere. As $r$ increases the line becomes weaker and the 
broad damping wings disappear (see Figure~\ref{valc_prd_r}). The $-Q/I$ 
profile of the line with $r=10^{-6}$ shows a maximum in both the line core 
and in the line wing (i.e., at about $\approx 15$ Doppler widths) as well as 
a minimum in the near wing (i.e., at about $\approx 4$ Doppler widths). 
As $r$ increases both the wing maximum and the wing minimum are shifted 
towards line center and decrease in 
magnitude, to eventually disappear (see the long-dashed line in 
Figure~\ref{valc_prd_r}). The line core polarization, however, initially 
decreases and becomes negative for $r=10^{-2}$ and then becomes positive again 
for $r=0.7$. This response of $-Q/I$ to the variations in $r$ can be understood 
using the mean anisotropy $\bar A_{\rm PRD}$ plotted in 
Figure~\ref{valc_prd_r}. It is worth 
to note that the shapes of the $-Q/I$ profiles that we obtain with PRD 
for $r$ values between $10^{-6}$ and $10^{-2}$ (see Figure~\ref{valc_prd_r}) 
are similar to those classified as M-signals by \citet{lbandeld09}, whereas 
the shape 
of the $-Q/I$ profile for $r=0.7$ is similar to that classified as S-signal by 
these authors. We recall from \citet{lbandeld09} that a $Q/I$ profile with 
a single peak at line center is classified as S-signal, while $Q/I$ profiles 
showing a polarization maximum in the wings, a decrease in amplitude 
approaching the line core, and eventually a narrow peak at line 
center are classified as M-signals. 

For $r=10^{-6}$ and $10^{-5}$ the $-Q/I$ profile is nearly proportional to 
$\bar A_{\rm PRD}$ in the frequency range $0\le x\le 5$ Doppler widths, so that 
Equation~(\ref{qoveri_eb}) is applicable. However for $x>5$, one cannot neglect 
the contribution of $r$ in $(S_{x})^0_0$ as was done to deduce 
Equation~(\ref{qoveri_eb}). In the presence of a background continuum, one can 
approximately write the 
emergent $Q/I$ profile as (still neglecting the thermal term $\epsilon B$) 
\begin{equation}
{Q\over I}(\tau_{x\mu}=0) \approx {3 \over 2\sqrt{2}}\,(\mu^2-1)\,W_2
{\phi_{x} \bar A_{\rm PRD}(\tau_{x\mu}=1) \over \phi_{x} + r\,B/(\bar J_x)^0_0}.
\label{qoveri_cont_eb}
\end{equation}
The term with $r$, in the denominator of Equation~(\ref{qoveri_cont_eb}), is 
responsible for the formation of a wing maximum in $-Q/I$. The fact that 
the introduction of $r$ gives rise to a wing maximum in $-Q/I$ was already 
noted for isothermal models by \citet{rs82}. This wing maximum is 
located at the frequency where the radiation field starts being 
influenced by the continuous absorption \citep[see][]{mf88}. In other words, at 
that frequency the line source function and the (unpolarized) 
continuum source function equally contribute to the total source 
function. For isothermal model atmosphere the position of this wing maximum 
can be estimated using the thermalization frequency given by \citet{hf80}. 

The frequency range where Equation~(\ref{qoveri_eb}) can be applied 
decreases as $r$ increases and can no longer be applied for $r> 10^{-2}$. 
On the other hand, when $r$ increases, the term containing $r$ in 
Equation~(\ref{qoveri_cont_eb}) starts 
influencing not only the wings but also the line core region, thereby 
greatly reducing the polarization in the wings to zero and confining the 
$-Q/I$ profile to the line core. It is worth to note that for $r \le 10^{-3}$ 
the mean anisotropy $\bar A_{\rm PRD}$ is positive in the line core and in the 
line wings showing that the radiation field is limb-darkened. For those 
frequencies $x$ whose $\tau_{x\mu}=1$ at the heights where we have the VALC 
temperature rise ($\approx 500$ -- $1000$\, km; see Figure~\ref{valcandmco}), 
the anisotropy is negative, showing that the radiation field is 
limb-brightened. This fact was already noted for schematic chromospheric-like 
models by \citet{mf88}. 

As $r$ increases, the anisotropy $\bar A_{\rm PRD}$
in the line core decreases and even becomes negative for $r=10^{-2}$. This 
is because the line core photons now start to originate in deeper layers 
of the atmosphere, where the radiation field for the line core photons tends to become more and more 
isotropic. For $r=10^{-2}$ a limb-brightened radiation field 
dominates to give a negative $\bar A_{\rm PRD}$ in the line core. However, 
for $r=0.7$, the anisotropy $\bar A_{\rm PRD}$ reaches 
a large positive value. Note that 
the corresponding line source function $S_{I}$ shows a larger departure 
from the Planck function, not only above the temperature minimum, but also in 
the photospheric range of heights 300 -- 500\,km (where actually the 
line core is formed; see Table~\ref{formation_height_valc}), differently 
from what happens with other values of $r$ (see the $S_{\rm I}$ panel of 
Figure~\ref{valc_prd_r}). In fact, by taking values of $r$ between $10^{-2}$ and 0.7, it can be shown that there 
is a gradual transition in the line-center anisotropy $\bar A_{\rm PRD}$ from 
negative values to a large positive value.

It is worth to note that $\bar A_{\rm PRD}$ in the wings attains a constant 
value, which is the same for all the values of $r$. However, the frequency 
distance from line center at which this 
constant value is reached decreases as $r$ increases. 
This can be understood by taking the example of $r=10^{-5}$. In this case 
we find that the Stokes parameters at all frequencies $x \ge 20$ are ``formed"  
at a height of 360\,km (in the sense that this is the height at which 
the monochromatic optical depth is unity). When we 
plot the total source function for all these frequencies we find 
that they nearly coincide with the Planck function and, moreover, that they are 
dominated by the continuum source function. Thus, the anisotropy at those 
frequencies is determined by the Planck function gradient at a height of 
360\,km. As $r$ increases the frequency value above which all the photons 
form at 360\,km progressively moves towards line center. 

A comparison of Figures~\ref{valc_prd_r} and \ref{valc_crd_r}, shows that 
CRD is a good approximation in the line core and particularly at 
line center. This is a well known result since the detailed work 
by \citet[][using the type I redistribution function]{dumontetal77} and 
\citet[][using the type II redistribution function]{mf87,mf88}. 
Furthermore, for sufficiently large values of $r$ (weak resonance lines) 
the CRD approximation can be safely used to model line profiles \citep{hf96}, 
like for instance the $Q/I$ profiles of the photospheric line of Sr\,{\sc i} 
at 4607\,\AA\ \citep[see][]{mf93}. This is because 
these lines do not have well-developed broad wings (unlike the strong 
resonance lines). Therefore transfer of photons in the line wings, where 
frequency coherent scattering can play a crucial role, is negligible 
\citep[see][]{mf93}. As a result, even in the PRD case, 
the polarization gets confined to the line core as in the CRD case. 

\subsubsection{Effect of Photon Destruction Probability Per Scattering 
$\epsilon$}
Figure~\ref{valc_prd_eps} shows the 
sensitivity of $I$, $-Q/I$, $\bar A_{\rm PRD} (\tau_{x\mu}=1)$ and 
$S_{\rm I}(x=0)$ at $\mu=0.11$ to 
changes in the $\epsilon$ value. The corresponding CRD case is shown in 
Figure~\ref{valc_crd_eps}. 

When $\epsilon$ increases the coupling between the 
radiation field and the Planck function becomes stronger. From 
Figure~\ref{valc_prd_eps} we clearly see that $S_{\rm I}\simeq B$ in a 
greater region of 
the atmosphere for larger values of $\epsilon$. As a result, the intensity 
increases. However, the anisotropy decreases in the line core. Moreover, for 
frequencies $x \simeq 2.5$ the absolute value of the anisotropy (which is 
here negative) increases, while for $x>7$ the anisotropy is independent 
of $\epsilon$. Note that $x\approx 7$ corresponds to the wing maximum  
in the $-Q/I$ profile. As already 
noted in Section~4.1.1, at this frequency the total source function starts 
being dominated 
by the unpolarized continuum, so that the anisotropy as well as $-Q/I$ are 
independent of $\epsilon$. A comparison with the corresponding CRD case 
in Figure~\ref{valc_crd_eps}, shows that the above discussion for the 
line core region can be used to understand the CRD profiles also. 

\subsubsection{Effect of the Damping Parameter $a$}
Figure~\ref{valc_prd_damp} shows 
the response of $I$, $-Q/I$, $\bar A_{\rm PRD}(\tau_{x\mu}=1)$ and 
$S_{\rm I}(x=0)$ at $\mu=0.11$ to variations of $a$. 
The corresponding CRD case is shown in Figure~\ref{valc_crd_damp}. 

The larger the damping parameter $a$ the broader the absorption profile. 
As a result the emergent $I$ as well as $-Q/I$ profiles broaden as $a$ 
increases. Furthermore, an increase of $a$ implies that the 
source function departs from the Planck 
function to a slightly larger extent (see Figures~\ref{valc_prd_damp} and 
\ref{valc_crd_damp}). 
In the CRD case the anisotropy decreases with increasing values of $a$ and the 
$-Q/I$ profile does the same. In the PRD case the anisotropy at line center 
does not show a large sensitivity to $a$, and thereby the same happens with 
$-Q/I$. On the contrary, for $x>3$, $\bar A_{\rm PRD}$ shows a large 
sensitivity to variations in $a$. In particular, the anisotropy increases in 
absolute value for frequencies that are formed in the temperature rise region 
of the VALC atmosphere. Furthermore, as $a$ increases the frequency at which 
$\bar A_{\rm PRD}$ reaches a constant value also 
increases. Consequently the negative dip (or wing minimum) and the wing 
maximum in $-Q/I$ are very sensitive to $a$. The position of the negative 
dip as well as that of the wing maximum shifts away from line center. 
Since the position of the wing maximum is determined by the frequency at which 
the total source function starts being dominated by the unpolarized continuum, 
it is clear that an increase in $a$ results in a shift towards larger 
frequencies of the wing maximum of $-Q/I$. 
It is worth to note that \citet{gs85} also shows the 
influence of $a$ on $I$ and $-Q/I$, though a detailed analysis is missing. 
Nevertheless, it is clear from our Figure~\ref{valc_prd_damp} that a 
height dependence of $a$ has important effects on the emergent polarization 
profiles, a point already stressed by \citet{gs85}. 

We point out that our assumption of a constant damping parameter $a$ 
throughout the atmosphere is actually in contradiction with the hypothesis 
of an exponentially stratified atmosphere. This is because, as $a$ is 
proportional to the density, it should also be exponentially stratified. 
We also point out that the damping parameter $a$ is proportional to the 
elastic collisional rate $\Gamma_{\rm E}$, that has been here assumed to be 
zero. Notwithstanding these inconsistencies, we considered worthwhile 
to perform our studies as if the damping parameter were an independent 
quantity.

\subsubsection{Effect of the Elastic Collisions}
Figure~\ref{valc_prd_gammae} shows the sensitivity of $I$, $-Q/I$, 
$\bar A_{\rm PRD}(\tau_{x\mu}=1)$ and $S_{\rm I}(x=0)$ at $\mu=0.11$ to the 
variations of $\Gamma_{\rm E}/\Gamma_{\rm R}$. The corresponding CRD 
case is shown in Figure~\ref{valc_crd_gammae}. The depolarizing elastic 
collision parameter is assumed to be $D^{(2)}=0.5\,\Gamma_{\rm E}$ 
\citep[e.g.,][]{jos94}. We recall that in the CRD case, the redistribution 
matrix is given by Equation~(\ref{crd_redist}). Clearly only $D^{(2)}$ is 
the relevant quantity for CRD. Therefore, for CRD varying 
$\Gamma_{\rm E}/\Gamma_{\rm R}$ is equivalent to varying 
$D^{(2)}/\Gamma_{\rm R}$ since we have assumed $D^{(2)}/\Gamma_{\rm R}=0.5\,
\Gamma_{\rm E}/\Gamma_{\rm R}$. We note that $\Gamma_{\rm E}/\Gamma_{\rm R}$ 
mainly operates in the line wing, while $D^{(2)}/\Gamma_{\rm R}$ operates 
in the line core \citep[see also][]{mf92,knn94}. 
In the solar atmosphere the elastic collisions $\Gamma_{\rm E}$ and $D^{(2)}$ 
increase with depth in the atmosphere, while they are negligible in the upper 
layers, namely in the upper chromosphere and transition region. Thus, depth 
dependent elastic collisions affect only the wings through 
$\Gamma_{\rm E}/\Gamma_{\rm R}$, while the line core is nearly unaffected as 
it is formed higher in the atmosphere where $D^{(2)}$ is negligible 
\citep[see also][]{mf92}. 

As expected, elastic collisions have little effect on the line source function 
$S_{\rm I}$. In the case of PRD, as $\Gamma_{\rm E}/\Gamma_{\rm R}$ operates 
in the line wings, there is here some sensitivity to this parameter 
in the intensity profile. However, for CRD, as 
$D^{(2)}/\Gamma_{\rm R}$ operates mainly in the line core, there is hardly 
any effect on the intensity. As $\Gamma_{\rm E}/\Gamma_{\rm R}$  or 
$D^{(2)}/\Gamma_{\rm R}$ increases, basically the mean anisotropy factors 
$\bar A_{\rm CRD}$ and $\bar A_{\rm PRD}$ decrease at all frequencies. In the 
case of PRD, with an increase in elastic collisions, the contribution of 
$g^{\rm III}_{xx^\prime}$ increases. Since $g^{\rm III}_{xx^\prime}$ 
behaves more like CRD, when its contribution increases $\bar A_{\rm PRD}$ in 
the wings decreases and it then gradually tends to zero. 

\subsubsection{Additional Studies}
Here we present some additional studies related to the PRD linear polarization 
profiles. From now on we do not present the corresponding CRD cases. 

In Figure~\ref{valc_prd_lambda} we show $I$, $-Q/I$, 
$\bar A_{\rm PRD}(\tau_{x\mu}=1)$ and $S_{\rm I}(x=0)$ at $\mu=0.11$ for two 
different spectral lines. Clearly, the gradient of the Planck function as well 
as of $S_{\rm I}$ are larger at shorter wavelengths (e.g., for 
$\lambda=3933$\,\AA\ compared to those for $\lambda=5000$\,\AA), particularly 
in the photosphere. Therefore, when we move to the blue part of the spectrum 
the anisotropy increases particularly in the wings, so that the polarization 
also increases there. 

Figure~\ref{valct_prd_r} shows $I$, $-Q/I$, $\bar A_{\rm PRD}(\tau_{x\mu}=1)$ 
and the line source function $S_{\rm I}(x=0)$ at $\mu=0.11$ for lines formed 
in the transition region of the VALC model. As $r$ decreases, 
the line is formed higher in the atmosphere (see 
Table~\ref{formation_height_valc}) where it encounters a steep 
temperature rise, so that the radiation field tends towards becoming 
limb-brightened. As a result, the anisotropy decreases in the line core, while 
it increases in absolute value at the negative dip ($3 \le x \le 7$). Moreover,
as $r$ decreases Equation~(\ref{qoveri_eb}) can now be applied to a larger 
frequency domain and hence the wing maximum in $-Q/I$, where the unpolarized 
continuum source function starts to dominate, now shifts towards larger 
frequencies. As the spectral line is formed in the region of the steep 
temperature rise, the intensity shows a typical emission profile, increasingly 
self-absorbed. 

Figure~\ref{valct_prd_w2} shows the effect of $W_2$ on $-Q/I$ 
and $\bar A(\tau_{x\mu}=1)$ at $\mu=0.11$, for the VALC 
model including its transition region and corona. Since $W_2$ 
has only marginal effects on the intensity and the line source function 
$S_{\rm I}$, we do not show them here. As expected, when $W_2$ decreases 
the polarization decreases everywhere. However, it is interesting to note 
that, in general, we expect a linear dependence on $W_2$ (since $(S_{lx})^2_0$ 
directly depends on $W_2$, see Equation~(\ref{slk0})). In other words, as 
$W_2$ is reduced to half, we expect the polarization to be also reduced to 
half its value throughout the profile when compared to the corresponding 
$W_2=1$ case. Instead, we see that for $x<3$, $-Q/I(W_2=0.5)$ is nearly 
$0.5\times -Q/I(W_2=1)$. But for $x>3$ there is no linear dependence on $W_2$. 
This can also be seen in the mean anisotropy $\bar A_{\rm PRD}$, which does not 
contain the $W_2$ factor in its definition (see 
Equations~(\ref{jbark0})--(\ref{j20}) and (\ref{anisotropy_factor})). If there 
were a linear dependence on $W_2$, the mean anisotropy $\bar A_{\rm PRD}$ 
should have been exactly the same for both $W_2=0.5$ and $W_2=1$. Instead, we 
see considerable difference for $x>3$. This difference can be attributed to the 
contribution of $Q$ to $(\bar J_x)^2_0$, as $I$ is only marginally affected 
by the value of $W_2$. Thus, we conclude that the scattering polarization 
depends non-linearly on $W_2$. This is true even for the CRD case. 
This fact was also noted by \citet{josandslg76}. 

\subsection{Results of Calculations in the MCO Model}
Here we study the effect of a relatively cool lower chromosphere 
on the emergent $Q/I$. As the sensitivity to parameters like 
$r$, $\epsilon$ and $a$ is already quite well understood for the VALC 
model, here we only present\,: (1) the comparison 
between the line profiles of VALC and MCO models, and (2) the influence of 
the elastic collisional rates $\Gamma_{\rm E}/\Gamma_{\rm R}$ and 
$D^{(2)}/\Gamma_{\rm R}$. We do not show the corresponding CRD case 
as the main effects are confined to the line core. 

Figure~\ref{mco_vs_valc_prd_standard} shows the effect of the thermal structure 
of the atmosphere on the polarized line formation. From Figure~\ref{valcandmco} 
we see that the temperature minimum is shifted from the height 
$z=500$\,km in the VALC model 
to 1000\,km in the MCO model. Thus, monochromatic radiation formed in the 
region 500 -- 1000\,km in the VALC model is limb-brightened, while it is  
limb-darkened in the MCO model (as there is no temperature rise in that 
region). As the region of temperature rise for the MCO model is located between 
1000 and 2000\,km, where the line core is formed, we find that the anisotropy 
in the line core is smaller for the MCO model than for the VALC model. The 
temperature rise tends to make the radiation field more 
limb-brightened. Finally, in the wings the anisotropy is smaller compared to 
that of the VALC model. This is because frequencies $x \ge 4$ are formed 
in the region 360 -- 500\,km, where the temperature gradient is larger for 
VALC than for MCO. 

Figure~\ref{mco_prd_gammae} shows the effect of varying the elastic collisional 
rate $\Gamma_{\rm E}/\Gamma_{\rm R}$ for the MCO model. Effects of varying 
$\Gamma_{\rm E}/\Gamma_{\rm R}$ or $D^{(2)}/\Gamma_{\rm R}$ for MCO model 
is more or less similar to that for the VALC model. We note that 
when $\Gamma_{\rm E}/\Gamma_{\rm R}$ increases, 
the differences in the anisotropy as well as in the polarization obtained 
from the MCO and VALC models decrease 
(for example compare the long-dashed lines in 
Figures~\ref{valc_prd_gammae} and \ref{mco_prd_gammae}). 

\section{Conclusions}
The interpretation of the second solar spectrum is an exciting and tough 
challenge in Solar Physics. To achieve this it is essential to solve the 
polarized radiative transfer equation taking into account various physical 
phenomena. In the case of strong resonance lines one of the important physical 
ingredients required for the modeling of the linear polarization profiles
is the so-called partial frequency redistribution (PRD). In this paper, 
therefore, we have studied the scattering polarization profiles formed under 
PRD. We use the angle-averaged version of the collisional redistribution 
matrix \citep[see][]{dh88,vb97b}, which is valid for a 
two-level atom with unpolarized lower level. 
For the numerical method of solution we 
have applied the very efficient and accurate symmetric Gauss-Seidel method 
presented in \citet{samandjtb10}. 

Our main focus in this paper has been to study the influence of various 
atmospheric parameters (namely, $\epsilon$, $r$, $a$, 
$\Gamma_{\rm E}/\Gamma_{\rm R}$, $D^{(2)}/\Gamma_{\rm R}$ and the temperature 
structure of the atmosphere) on the linear polarization profiles. 
Interestingly, for strong lines the $-Q/I$ profiles that we obtain with PRD, 
for a wide range of the various parameters, are of the shape that has been 
classified as M-signals by \citet{lbandeld09}. We recall that M-signals 
represent $Q/I$ profiles with a three-peak structure. Therefore, we find 
that for strong resonance lines the three-peak structure for $Q/I$ is a 
common feature for various combinations of the other relevant parameters, 
thereby confirming previous results either obtained using more crude model 
atmospheres (isothermal atmospheres) or with less sophisticated modeling. 

We interpret the linear polarization profiles using the Eddington-Barbier 
relation and the anisotropy at the atmospheric height where 
$\tau_{x\mu}=1$. To this end, as in the CRD 
case, we define a suitably frequency integrated or mean anisotropy factor 
$\bar A$ for PRD, that can be directly related to the emergent linear 
polarization. We have shown in great detail that 
the emergent scattering polarization is a mapping of the anisotropy at $\tau_{x\mu}=1$ 
\citep[cf.,][]{rs82,jtb01,ll04}. 
Our study shows that a height dependent damping parameter $a$ has 
significant influence on the emergent linear polarization profiles, 
particularly in the wings of strong resonance lines. In general, scattering 
polarization profiles are very sensitive to the thermal structure of the 
atmosphere (see Figure~\ref{mco_vs_valc_prd_standard}). 
For the range of parameters considered in this paper, we find also that 
the dips in $Q/I$ are located at values of reduced frequency $x$ ranging 
from 2 to 5. Furthermore, the values of $Q/I$ at these dips (including 
their sign) seem to be very sensitive to the thermal structure and to 
the various atmospheric parameters. 

\begin{figure*}
\plotone{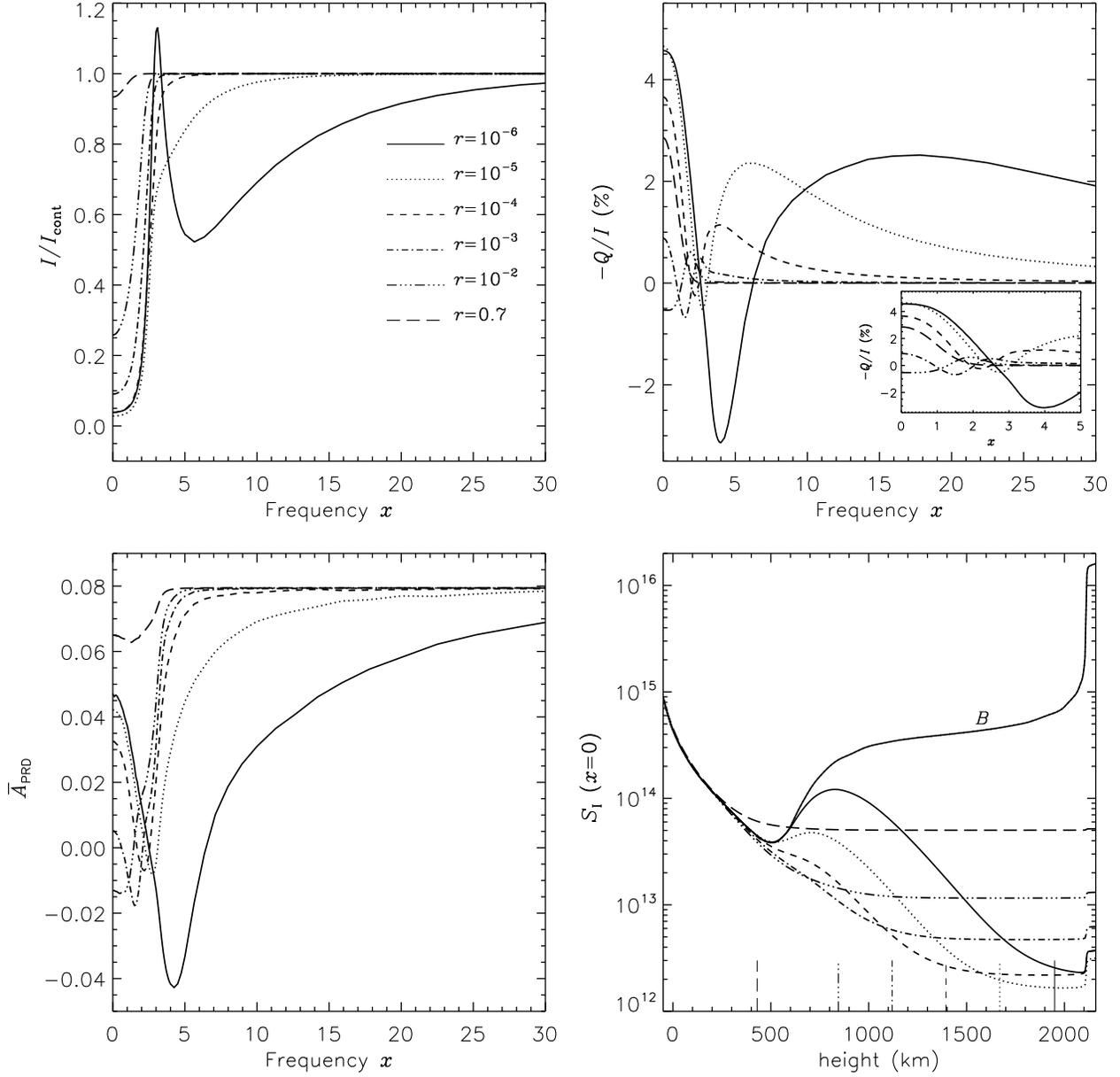}
\caption{Sensitivity of the PRD solutions in the VALC model to the line 
strength parameter $r$. 
The different line types are the following. Solid\,: $r=10^{-6}$; 
dotted\,: $r=10^{-5}$; dashed\,: $r=10^{-4}$; dot-dashed\,: $r=10^{-3}$; 
dash-triple-dotted\,: $r=10^{-2}$; long-dashed lines\,: $r=0.7$. Other 
model parameters are $\lambda=5000$\,\AA,\ $\epsilon=10^{-4}$, $a=10^{-3}$ and 
$\Gamma_{\rm E}/\Gamma_{\rm R}=D^{(2)}/\Gamma_{\rm R}=0$. The 
top solid line in the $S_{\rm I}$ panel is the Planck function $B$ for the temperature 
stratification of the VALC model. The vertical lines in the same panel 
show the height at which $\tau_{0\mu}=1$ for a LOS with $\mu=0.11$. The inset 
in the $-Q/I$ panel shows 
the line core region in more detail. The symbol $I_{\rm cont}$ denotes the 
intensity at very large distance from line center. 
}
\label{valc_prd_r}
\end{figure*}

\begin{figure*}
\plotone{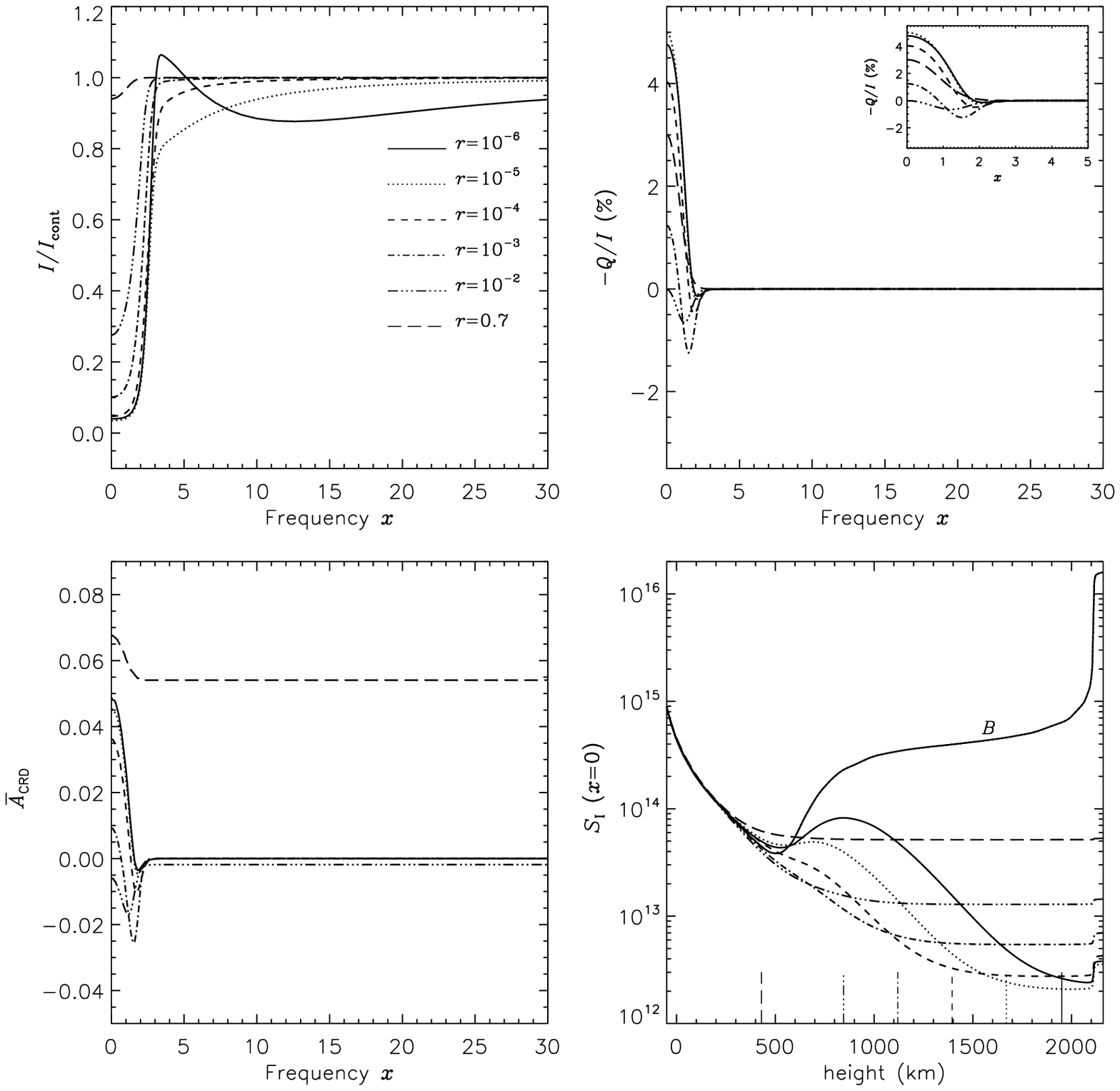}
\caption{Same as Figure~\ref{valc_prd_r}, but for CRD. The line types and the 
model parameters are exactly the same as in Figure~\ref{valc_prd_r}. 
}
\label{valc_crd_r}
\end{figure*}

\begin{figure*}
\plotone{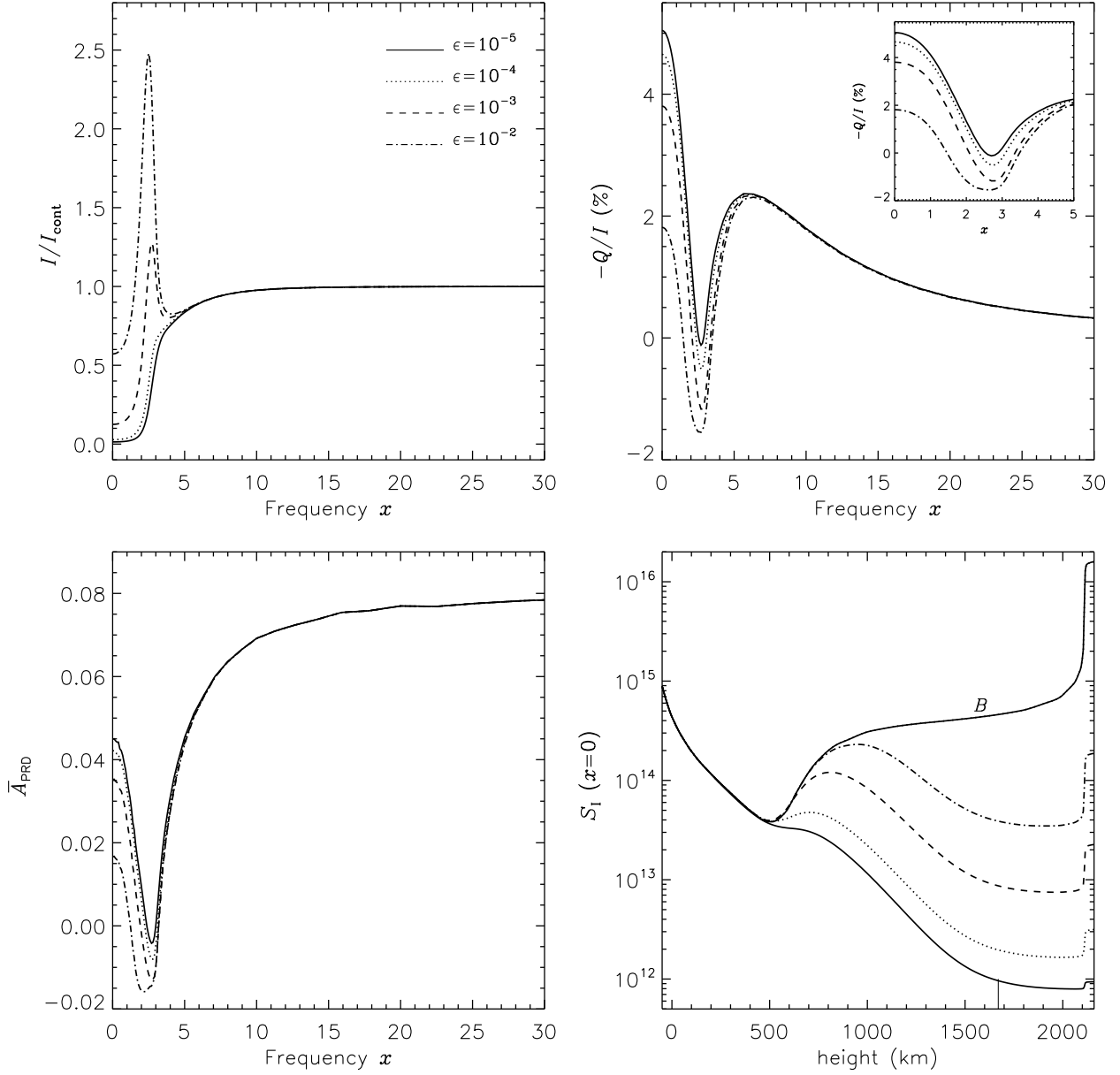}
\caption{The effect of $\epsilon$ on PRD solutions in the VALC model. 
The different line types are the following. Solid\,: $\epsilon=10^{-5}$; 
dotted\,: $\epsilon=10^{-4}$; dashed\,: $\epsilon=10^{-3}$; dot-dashed\,: 
$\epsilon=10^{-2}$. Other model parameters 
are $\lambda=5000$\,\AA,\ $r=10^{-5}$, $a=10^{-3}$ and 
$\Gamma_{\rm E}/\Gamma_{\rm R}=D^{(2)}/\Gamma_{\rm R}=0$. 
}
\label{valc_prd_eps}
\end{figure*}

\begin{figure*}
\plotone{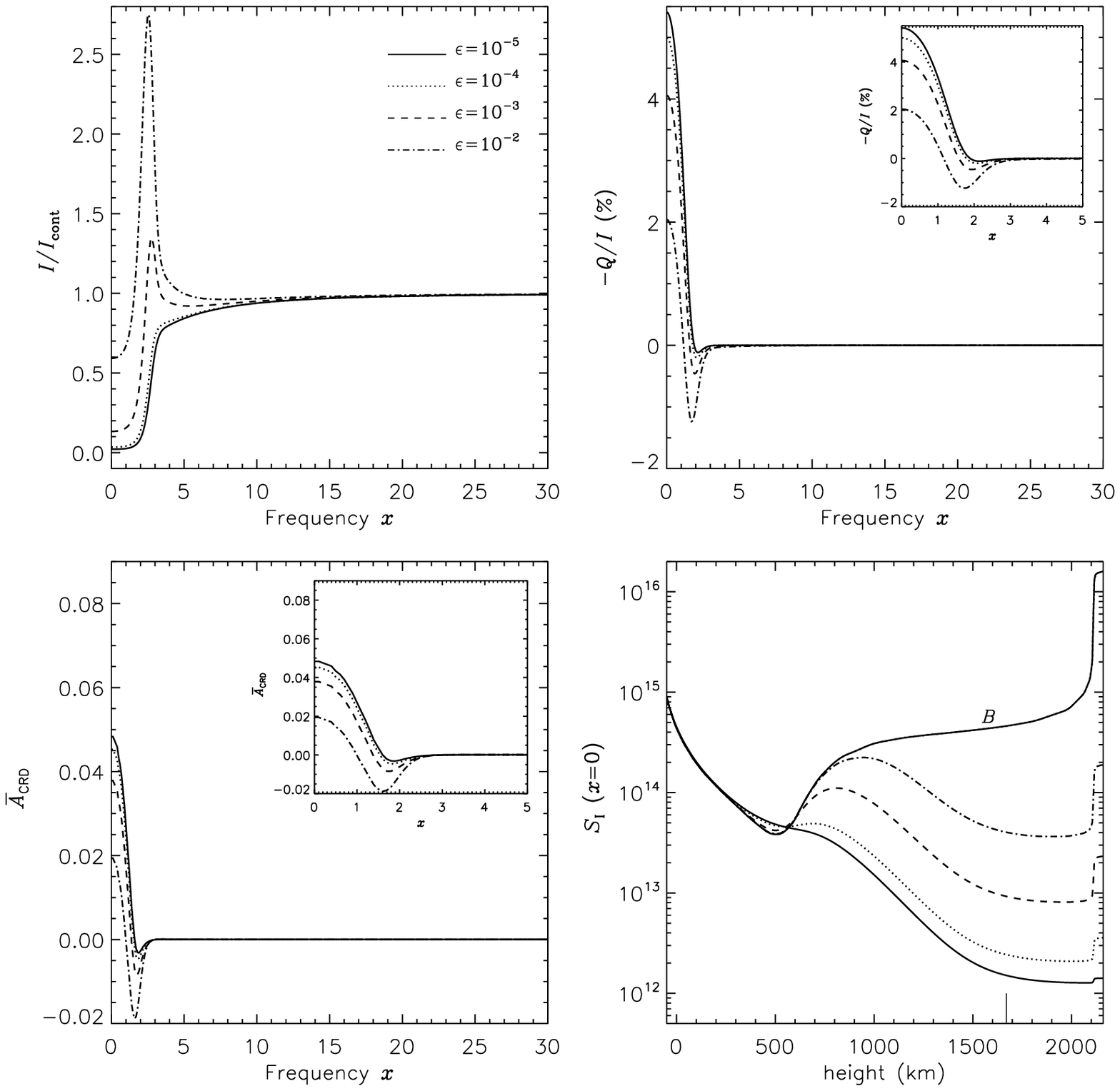}
\caption{Same as Figure~\ref{valc_prd_eps}, but for CRD. The line types and the 
model parameters are exactly the same as in Figure~\ref{valc_prd_eps}. 
}
\label{valc_crd_eps}
\end{figure*}

\begin{figure*}
\plotone{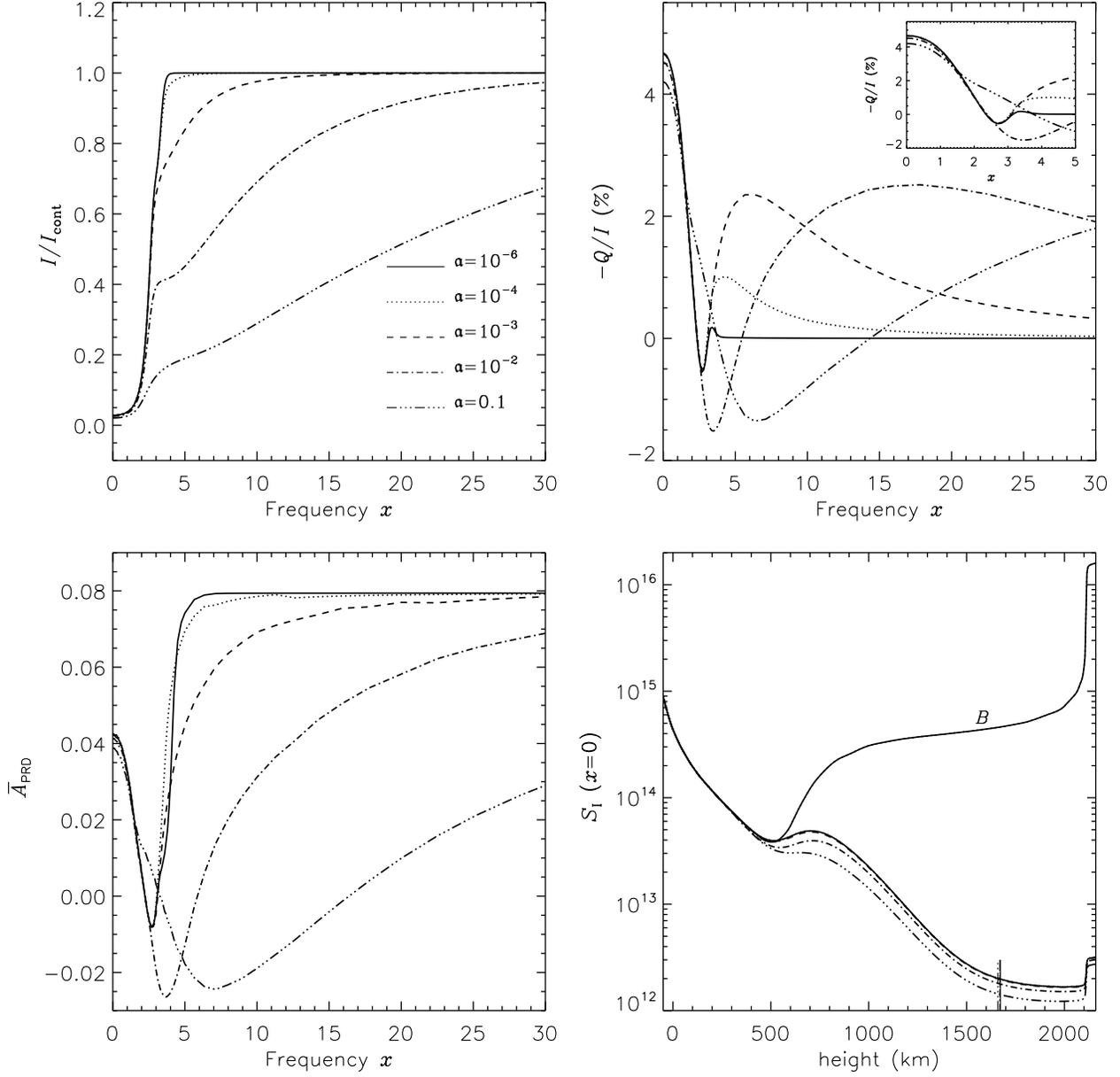}
\caption{Sensitivity of the PRD solutions in the VALC model to variations 
in the damping parameter $a$. The different line types are the following. 
Solid\,: $a=10^{-6}$; dotted\,: $a=10^{-4}$; dashed\,: $a=10^{-3}$; 
dot-dashed\,: $a=10^{-2}$; dash-triple-dotted\,: $a=0.1$. Other model 
parameters are $\lambda=5000$\,\AA,\ $\epsilon=10^{-4}$, $r=10^{-5}$, and 
$\Gamma_{\rm E}/\Gamma_{\rm R}=D^{(2)}/\Gamma_{\rm R}=0$. 
}
\label{valc_prd_damp}
\end{figure*}

\begin{figure*}
\plotone{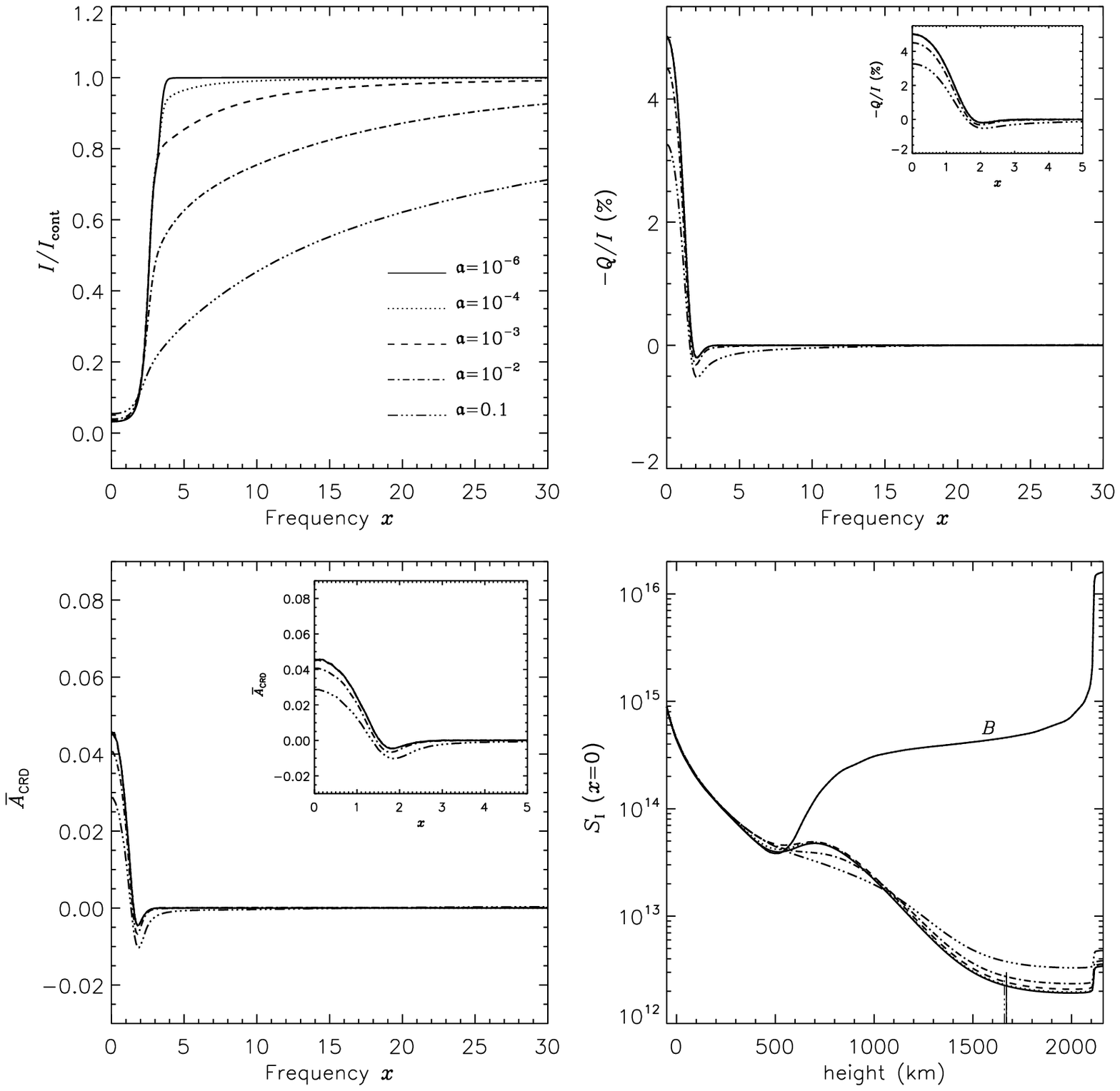}
\caption{Same as Figure~\ref{valc_prd_damp}, but for CRD. The line types and 
the model parameters are exactly the same as in Figure~\ref{valc_prd_damp}. 
}
\label{valc_crd_damp}
\end{figure*}

\begin{figure*}
\plotone{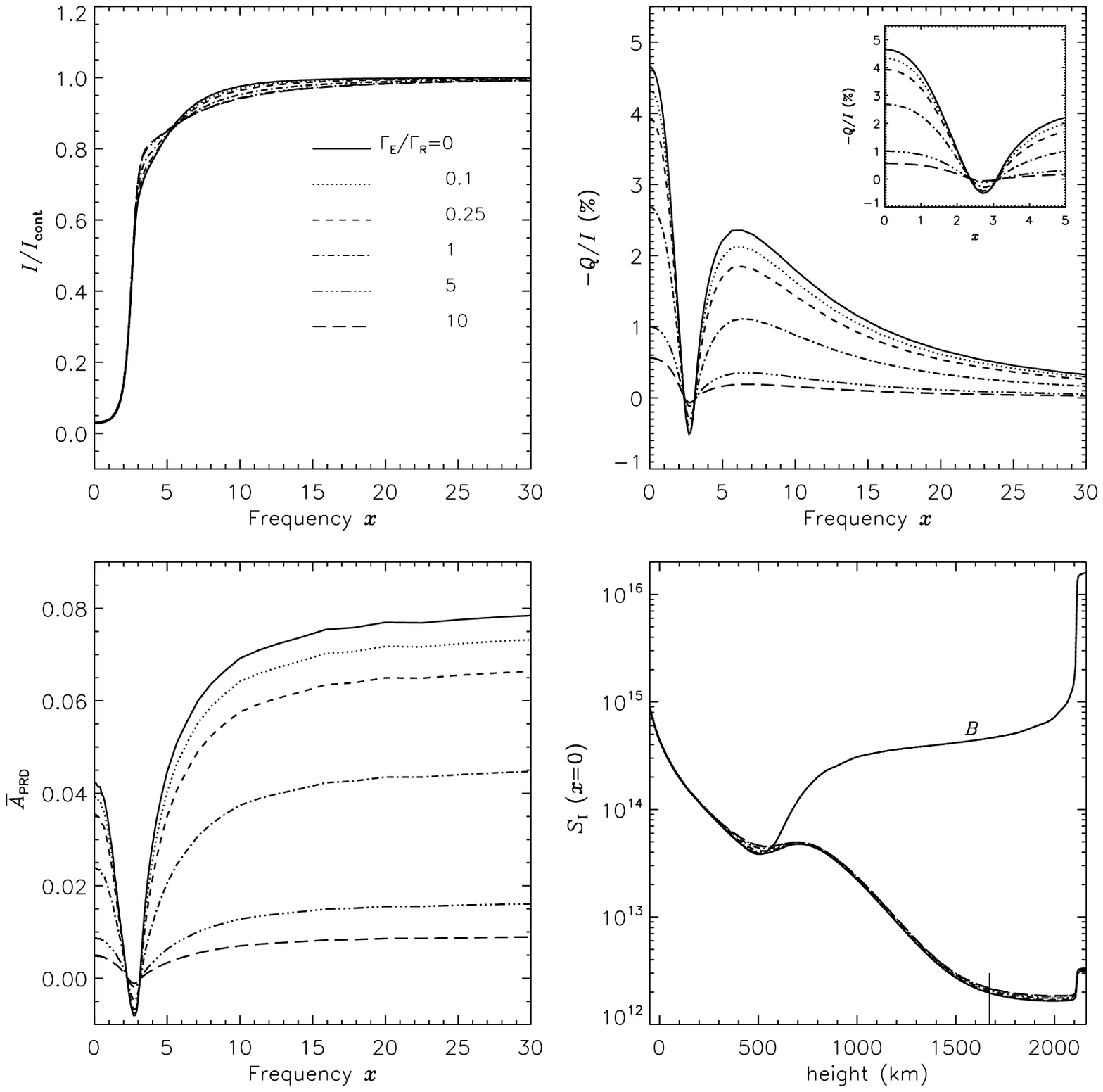}
\caption{The effect of the elastic collisional rate 
$\Gamma_{\rm E}/\Gamma_{\rm R}$ on the PRD solutions 
in the VALC model. The different line types are the following. 
Solid\,: $\Gamma_{\rm E}/\Gamma_{\rm R}=0$; dotted\,: 
$\Gamma_{\rm E}/\Gamma_{\rm R}=0.1$; 
dashed\,: $\Gamma_{\rm E}/\Gamma_{\rm R}=0.25$; 
dot-dashed\,: $\Gamma_{\rm E}/\Gamma_{\rm R}=1$; 
dash-triple-dotted\,: $\Gamma_{\rm E}/\Gamma_{\rm R}=5$;
long-dashed\,: $\Gamma_{\rm E}/\Gamma_{\rm R}=10$. 
Other model parameters are $\lambda=5000$\,\AA,\ 
$\epsilon=10^{-4}$, $r=10^{-5}$, $a=10^{-3}$ and 
$D^{(2)}/\Gamma_{\rm R}=0.5\,\Gamma_{\rm E}/\Gamma_{\rm R}$. 
}
\label{valc_prd_gammae}
\end{figure*}

\begin{figure*}
\plotone{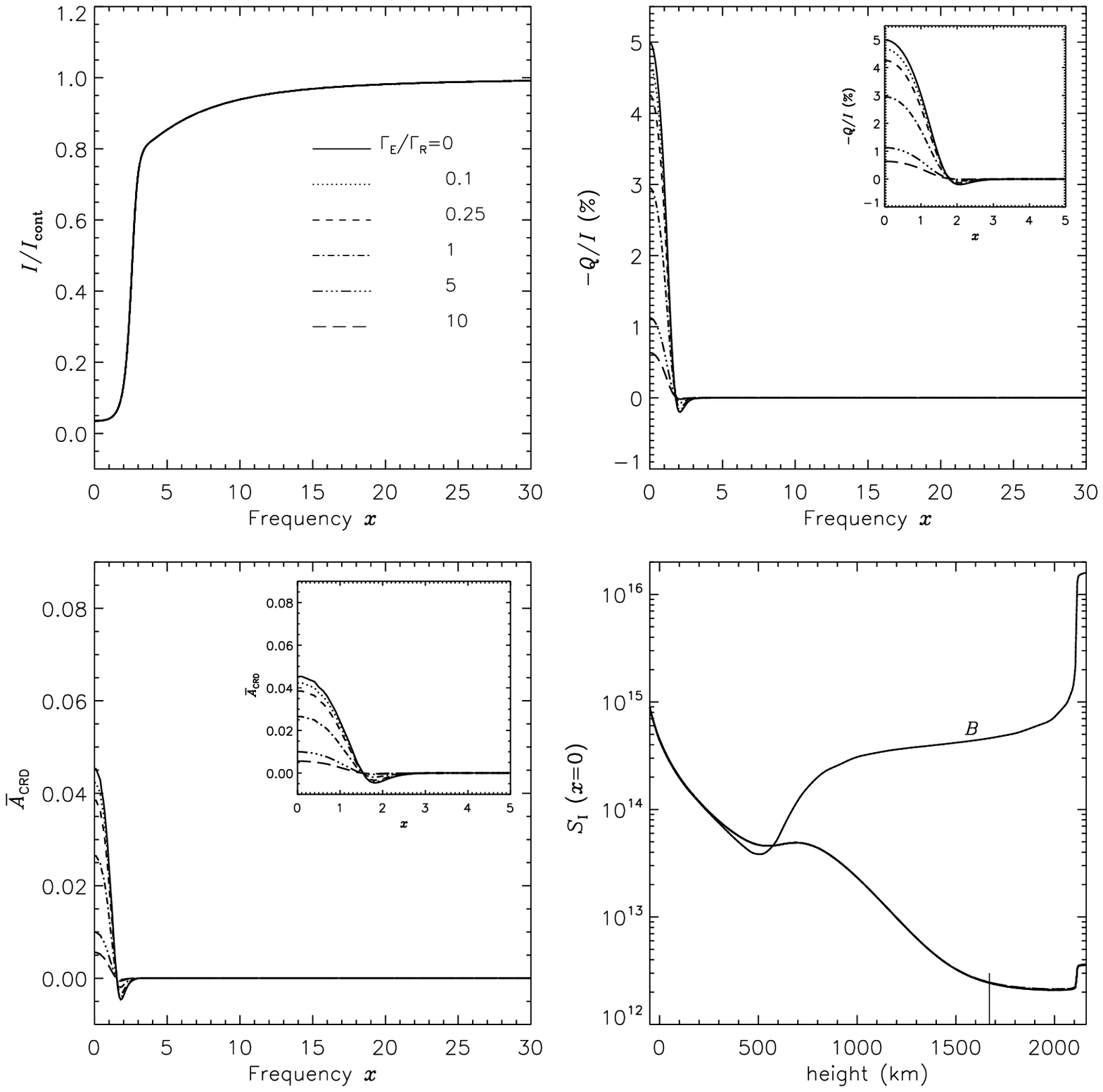}
\caption{Same as Figure~\ref{valc_prd_gammae}, but for CRD. The line types and 
the model parameters are exactly the same as in Figure~\ref{valc_prd_gammae}. 
}
\label{valc_crd_gammae}
\end{figure*}

\begin{figure*}
\plotone{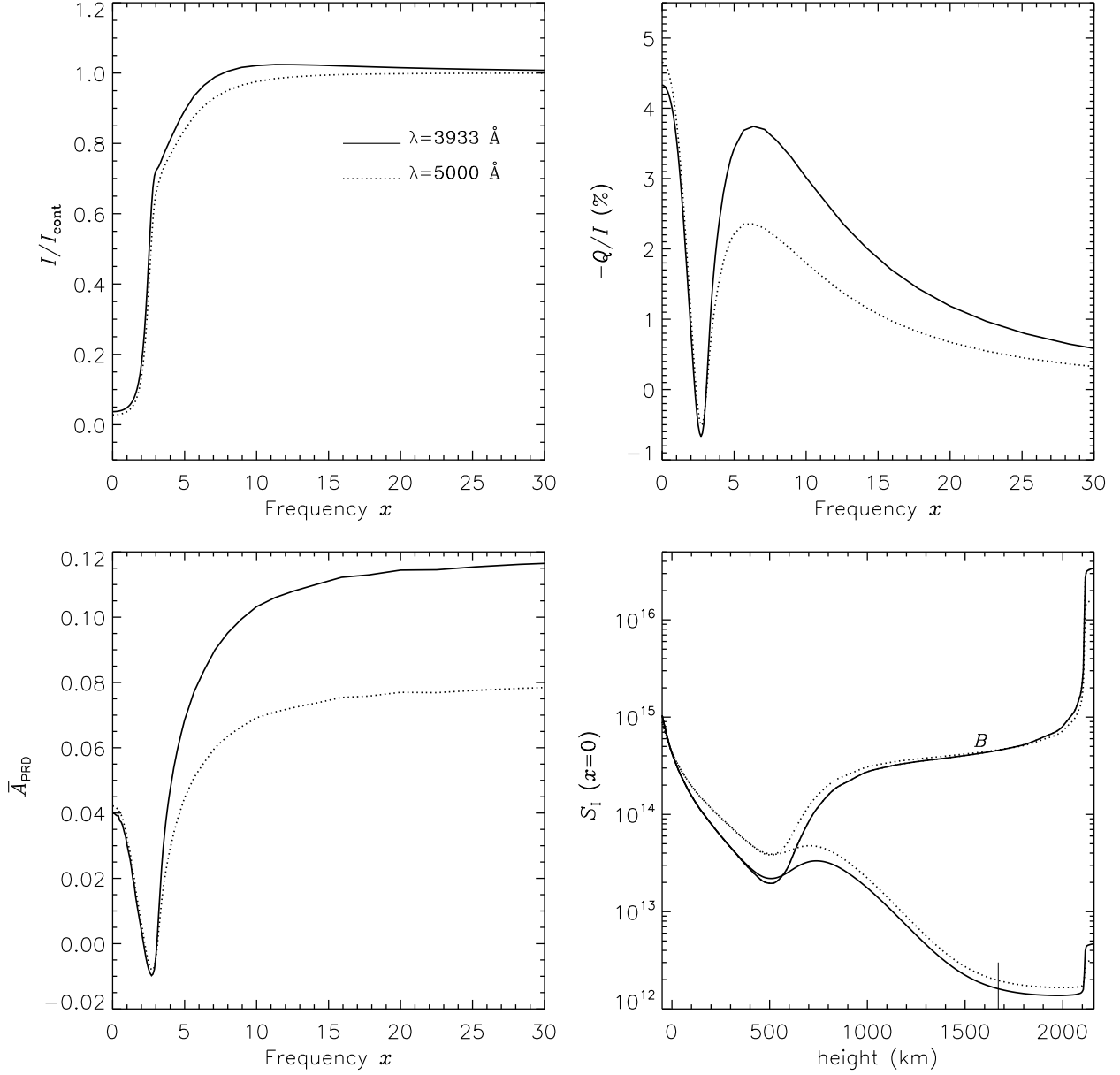}
\caption{PRD solutions in the VALC model for two spectral 
lines with different wavelengths. Solid line\,: 
$\lambda=3933$\,\AA;\ dotted line\,: $\lambda=5000$\,\AA.\ 
Other model parameters are 
$\epsilon=10^{-4}$, $r=10^{-5}$, $a=10^{-3}$ and 
$\Gamma_{\rm E}/\Gamma_{\rm R}=D^{(2)}/\Gamma_{\rm R}=0$. 
}
\label{valc_prd_lambda}
\end{figure*}

\begin{figure*}
\plotone{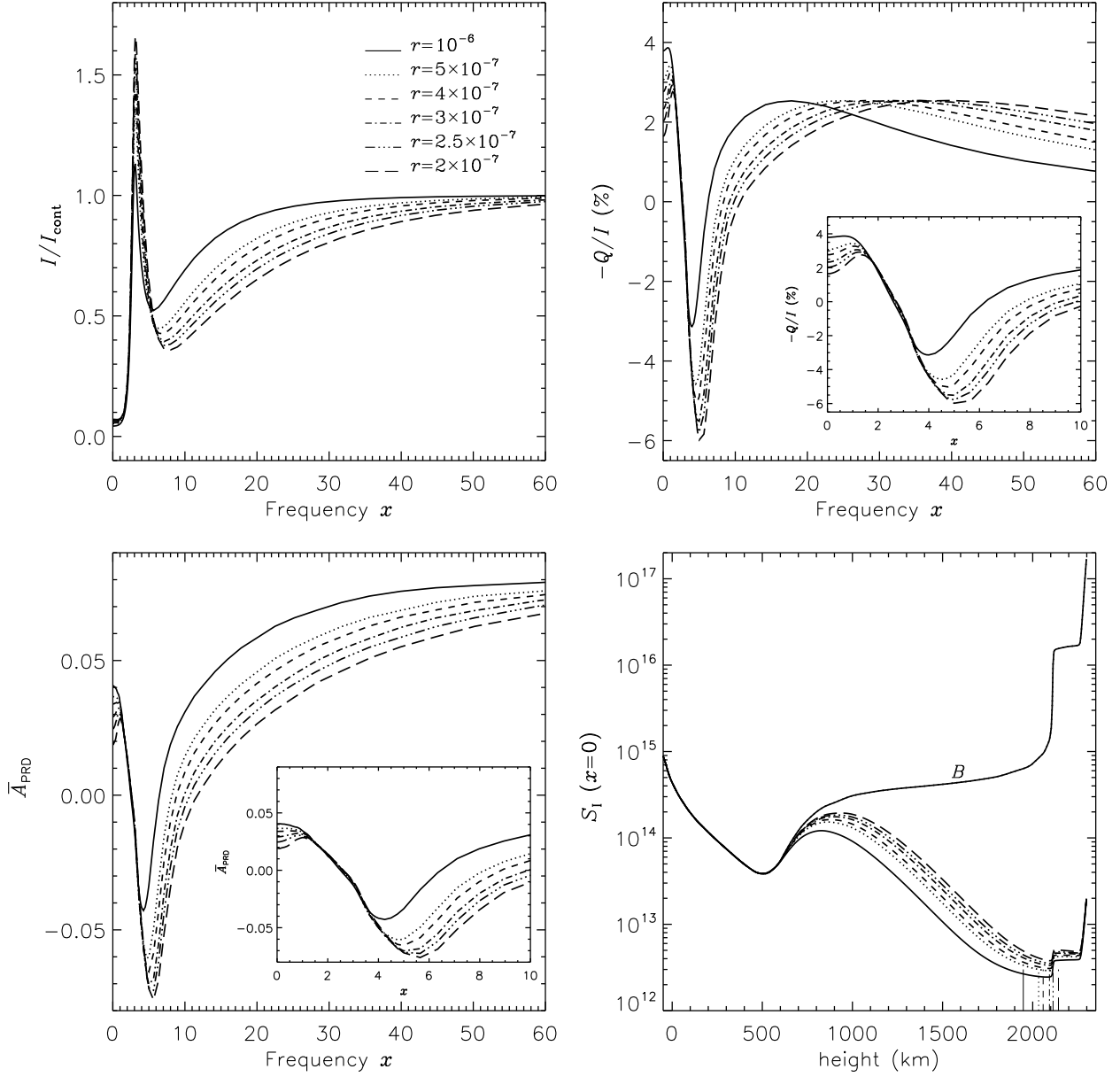}
\caption{PRD solutions in the VALC model for spectral lines of the upper 
chromosphere and transition region. 
The different line types are the following. Solid\,: $r=10^{-6}$; 
dotted\,: $r=5\times 10^{-7}$; 
dashed\,: $r=4\times 10^{-7}$; dot-dashed\,: $r=3\times 10^{-7}$; 
dash-triple-dotted\,: $r=2.5\times 10^{-7}$; long-dashed lines\,: 
$r=2\times 10^{-7}$. Other model parameters are $\lambda=5000$\,\AA,\ 
$\epsilon=10^{-4}$, $a=10^{-3}$ and 
$\Gamma_{\rm E}/\Gamma_{\rm R}=D^{(2)}/\Gamma_{\rm R}=0$. 
}
\label{valct_prd_r}
\end{figure*}

\begin{figure*}
\plotone{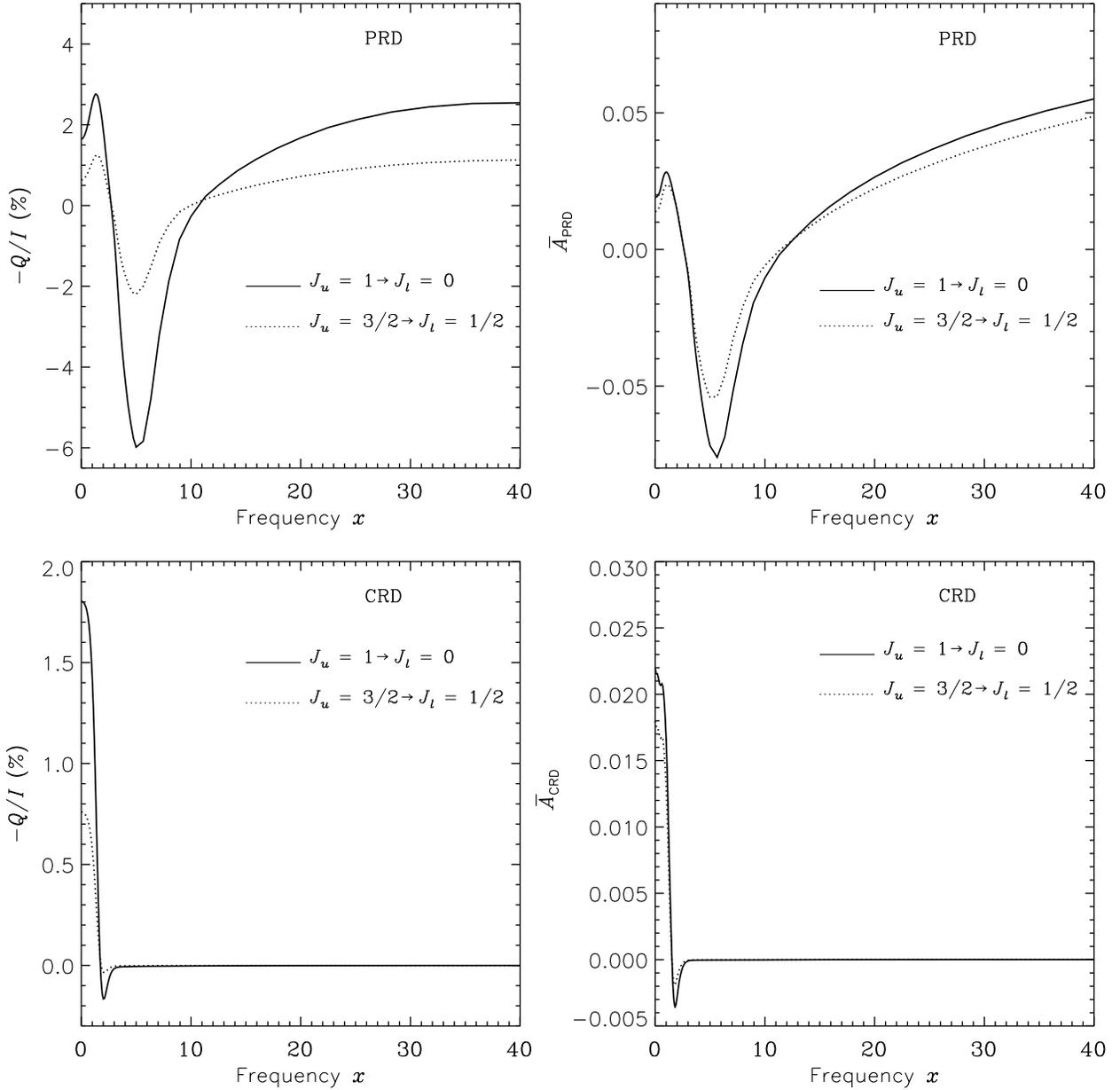}
\caption{The effect of the $W_2$ factor on the PRD (top panels) and CRD 
(bottom panels) solutions in the VALC 
model for a strong line with $r=2\times 10^{-7}$. 
Solid line\,: $W_2=1$ (which corresponds to a line transition with 
$J_l=0$ and $J_u=1$); dotted line\,: $W_2=0.5$ (which corresponds to a 
line transition with $J_l=1/2$ and $J_u=3/2$). 
Other model parameters are $\lambda=5000$\,\AA,\ 
$\epsilon=10^{-4}$, $a=10^{-3}$ and 
$\Gamma_{\rm E}/\Gamma_{\rm R}=D^{(2)}/\Gamma_{\rm R}=0$. 
}
\label{valct_prd_w2}
\end{figure*}

\begin{figure*}
\plotone{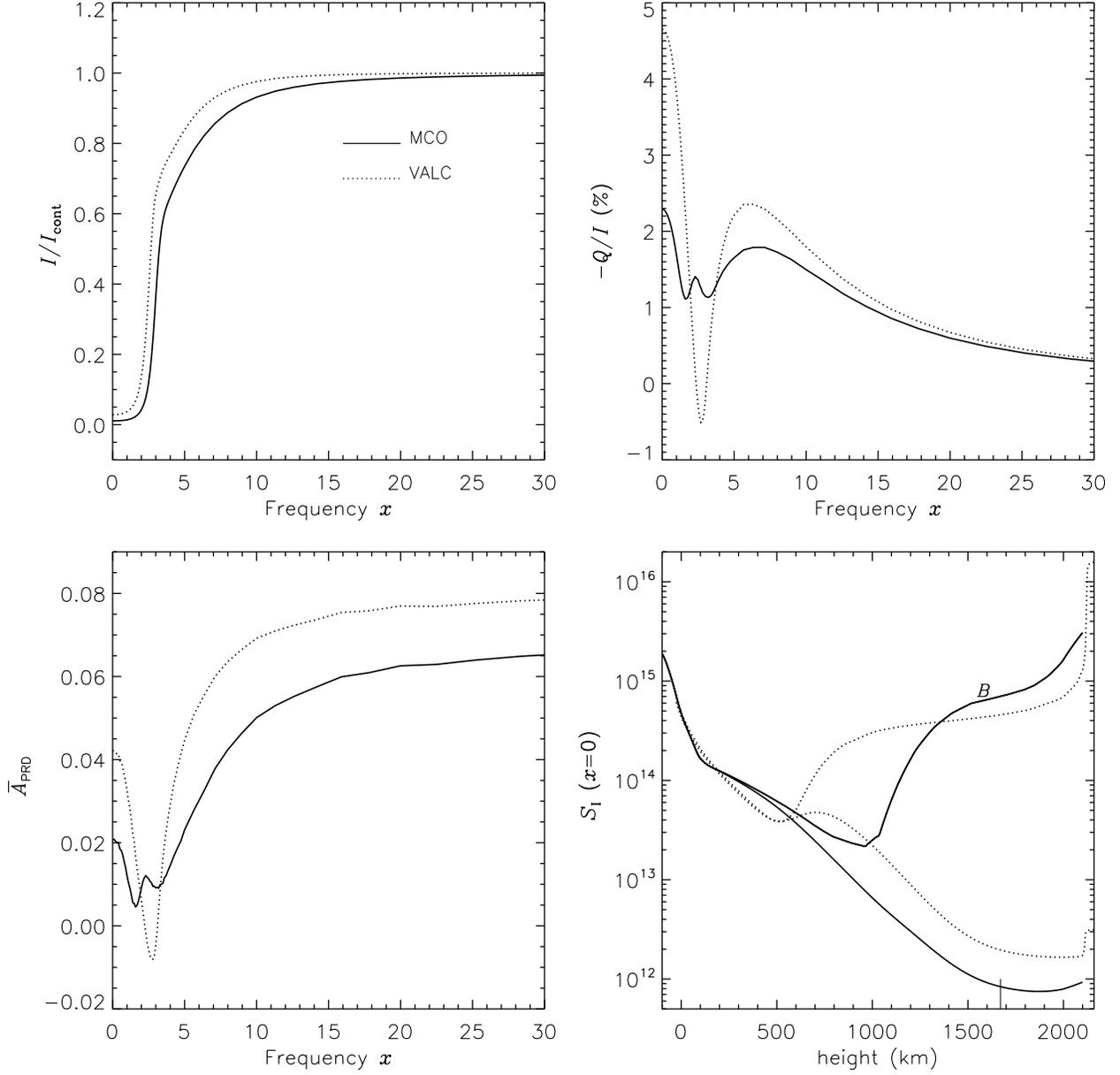}
\caption{The effect of the atmospheric thermal structure on the PRD 
solutions. Solid lines\,: MCO model, dotted lines\,: VALC model. 
Other model parameters are $\lambda=5000$\,\AA,\ 
$\epsilon=10^{-4}$, $r=10^{-5}$, $a=10^{-3}$ and 
$\Gamma_{\rm E}/\Gamma_{\rm R}=D^{(2)}/\Gamma_{\rm R}=0$. 
}
\label{mco_vs_valc_prd_standard}
\end{figure*}

\begin{figure*}
\plotone{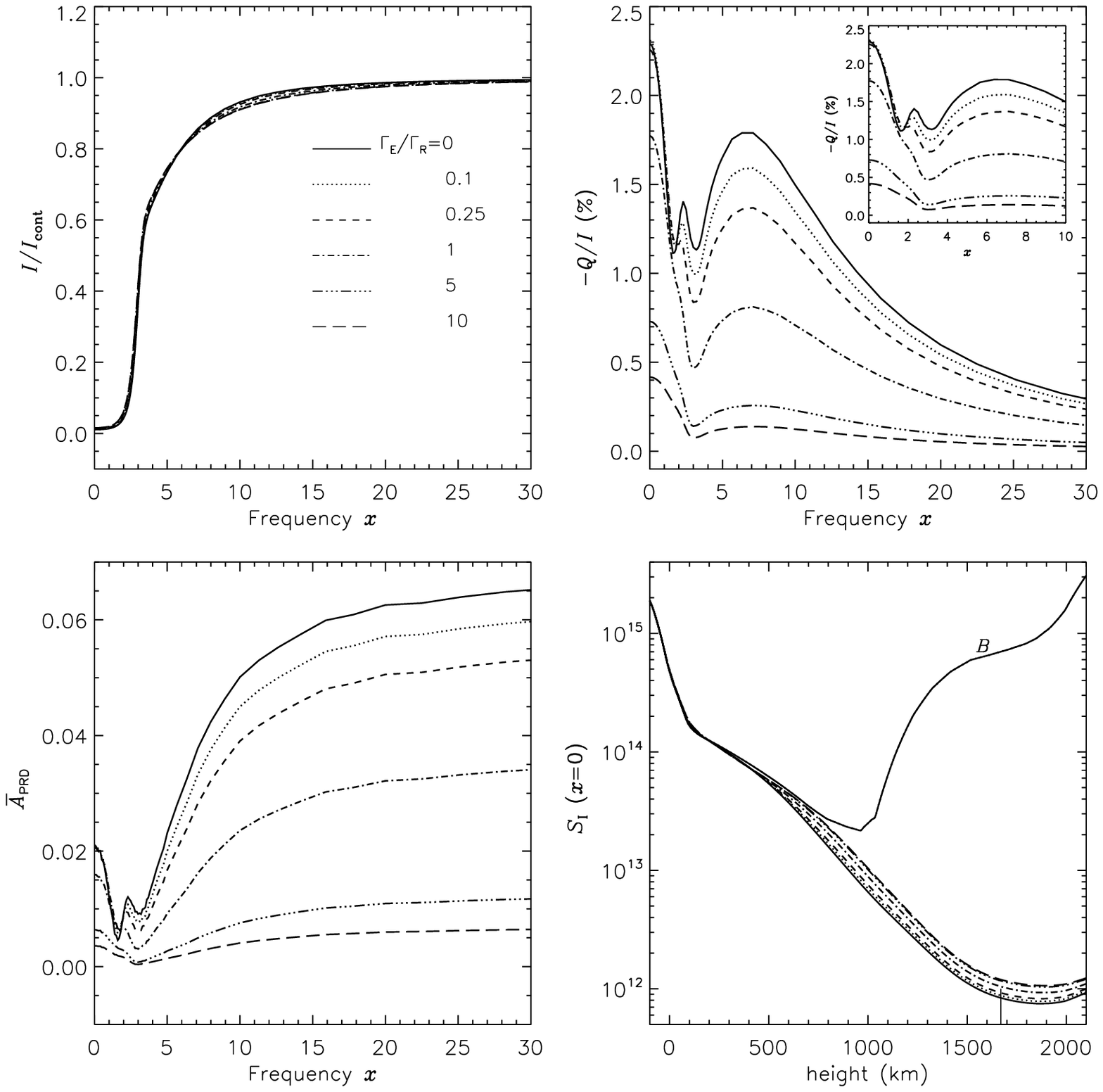}
\caption{The effect of the elastic collisional rate 
$\Gamma_{\rm E}/\Gamma_{\rm R}$ on the PRD solutions 
in the MCO model. The different line types are the following. Solid\,: 
$\Gamma_{\rm E}/\Gamma_{\rm R}=0$; dotted\,: 
$\Gamma_{\rm E}/\Gamma_{\rm R}=0.1$; 
dashed\,: $\Gamma_{\rm E}/\Gamma_{\rm R}=0.25$; 
dot-dashed\,: $\Gamma_{\rm E}/\Gamma_{\rm R}=1$; 
dash-triple-dotted\,: $\Gamma_{\rm E}/\Gamma_{\rm R}=5$;
long-dashed lines\,: $\Gamma_{\rm E}/\Gamma_{\rm R}=10$. 
Other model parameters are $\lambda=5000$\,\AA,\ 
$\epsilon=10^{-4}$, $r=10^{-5}$, $a=10^{-3}$ and 
$D^{(2)}/\Gamma_{\rm R}=0.5\,\Gamma_{\rm E}/\Gamma_{\rm R}$. 
}
\label{mco_prd_gammae}
\end{figure*}

The results of our investigation emphasize that PRD is required to model the 
linear polarization of strong resonance  
lines like Ca\,{\sc ii} K and Mg\,{\sc ii} k (compare Figures~\ref{valc_prd_r} 
and \ref{valc_crd_r}, see also Figure~\ref{valct_prd_r}), while CRD is a good 
approximation to model weak resonance lines like Sr\,{\sc i} 4607\,\AA\ 
\citep[see also][and compare the long-dashed lines of our Figures~\ref{valc_prd_r} and 
\ref{valc_crd_r}]{mf93,hf96}. A strong resonance line is characterized 
by broad damping wings, where frequency coherent scattering can play a 
crucial role \citep[e.g.,][]{hf96}. Weak resonance lines lack such 
damping wings and, therefore, the corresponding linear polarization, even 
computed in PRD, gets confined to the line core, with a single peak at 
line center, similarly to what is obtained when assuming CRD. 
In order to quantify the difference between the linear 
polarization profiles computed with CRD and PRD for the case of weak resonance lines, we define the following two 
quantities\,:
\begin{equation}
\delta\left[{Q\over I}(x=0)\right] = {|(Q/I)_{\rm PRD}(x=0)-
(Q/I)_{\rm CRD}(x=0)| \over |(Q/I)_{\rm PRD}(x=0)|},
\label{deltaqbyix0}
\end{equation}
which gives the relative difference in amplitude of $Q/I$ computed using CRD 
and PRD at line center, and 
\begin{equation}
\delta\left[{\rm HWHM}\right] = {|({\rm HWHM})_{\rm PRD}-
({\rm HWHM})_{\rm CRD}| \over |({\rm HWHM})_{\rm PRD}|},
\label{deltahwhm}
\end{equation}
which gives the relative difference in the half width at half maximum (HWHM) 
of the $Q/I$ profiles computed using CRD and PRD. Taking the particular case 
of $r=0.7$ (see long-dashed lines in Figures~\ref{valc_prd_r} and 
\ref{valc_crd_r}), we varied $\epsilon$, $a$, and 
$\Gamma_{\rm E}/\Gamma_{\rm R}$ as done in Figures~\ref{valc_prd_eps} -- 
\ref{valc_crd_gammae} for $r=10^{-5}$. We find that $\delta\left[(Q/I)(x=0)
\right]$ is around 5\,\% and $\delta[{\rm HWHM}]$ is about 10\,\% for 
the above-mentioned parametric study with fixed $r=0.7$. Clearly, the error 
introduced through the use of CRD for weak resonance lines is quite small. 
Therefore, we may conclude that CRD is a good approximation for weak resonance 
lines. 

Future investigations should extend the present work by considering
angle-dependent redistribution functions and the Hanle effect of deterministic magnetic fields.
On the one hand, radiative transfer calculations with angle-dependent redistribution functions are numerically 
very expensive \citep[e.g.,][]{mf88,knnetal02}. To assume the angle-averaged 
PRD function used here instead of the general angle-dependent function is justified only when 
polarization is neglected \citep[e.g.,][]{hf96}. Since resonance 
polarization is largely controlled by the anisotropy of the radiation field, 
the use of angle-averaged functions may not be as suitable for polarized 
transfer as for non-polarized transfer. Tests of this approximation performed 
by \citet{mf88} for purely coherent scattering in semi-infinite isothermal 
atmospheres show that the errors in the linear polarization are very small 
in the wings, while in the line core the polarization peak obtained with the 
angle-dependent type II redistribution function is slightly sharper than 
with $R_{\rm II,AA}$. However, such tests need to be revised considering
more realistic solar model atmospheres. On the other hand, polarized 
radiative transfer computations of the Hanle effect with angle-averaged PRD have 
been performed for isothermal model atmospheres by several researchers 
\citep[e.g.,][]{mf91,knnetal99,flurietal03,sametal08b}. Even though the Hanle effect 
is confined to the line core, there are considerable differences 
in the linear polarization profiles computed using CRD and angle-averaged 
PRD functions \citep[e.g.,][]{mf91,knnetal99}. Furthermore, 
\citet[][see also Sampoorna et al. 2008a]{knnetal02} have investigated the reliability of 
angle-averaged functions for the case of Hanle effect in isothermal 
atmospheres, showing that the Stokes $U$ parameter is relatively more 
sensitive than Stokes $Q$ regarding the use of angle-averaged 
versus angle-dependent functions. Therefore, to 
further assess the relative importance of PRD effects, we need to carry out 
detailed investigations similar to the one reported in this paper but including 
the Hanle effect of weak magnetic fields. Obviously, 
such radiative transfer problems are significantly more complicated that the one considered here, but
we think that the same numerical method presented in \citet{samandjtb10} can be suitably generalized 
to solve efficiently such more general PRD problems, both in 1D and 3D geometries. 

\acknowledgments
We are grateful to Luca Belluzzi and to the referee 
for carefully reviewing our paper. Thanks are also due to Dr. V. Bommier 
for providing a FORTRAN routine to compute the type III redistribution 
function and to Dr. K. N. Nagendra for valuable scientific discussions. 
Financial support by the Spanish Ministry of Science and Innovation 
through projects AYA2007-63881 (Solar Magnetism and High-Precision 
Spectropolarimetry) and CONSOLIDER INGENIO CSD2009-00038 (Molecular 
Astrophysics: The Herschel and Alma Era) is gratefully acknowledged.

\newpage
\begin{deluxetable}{ccc}
\tablewidth{0pt}
\tablecaption{The height in the VALC model atmosphere at which lines of 
different strength $r$ have $\tau_{0\mu}=1$. 
Other model parameters are 
$\lambda=5000$\,\AA,\ $\epsilon=10^{-4}$, $a=10^{-3}$ and 
$\Gamma_{\rm E}/\Gamma_{\rm R}=D^{(2)}/\Gamma_{\rm R}=0$. 
\label{formation_height_valc}}
\tablehead{
\colhead{$r$} & \colhead{$z(\tau_{x\mu}=1)$ in km} & 
\colhead{$z(\tau_{x\mu}=1)$ in km}\\
\colhead{} & \colhead{$x=0$ and $\mu=0.11$} & 
\colhead{$x=0$ and $\mu=1$}
}
\startdata
$2\times 10^{-7}$ & 2148 & 1878 \\
$2.5\times 10^{-7}$ & 2118 & 1848 \\
$3\times 10^{-7}$ & 2088 & 1833 \\
$4\times 10^{-7}$ & 2058 & 1788 \\
$5\times 10^{-7}$ & 2028 & 1773 \\
$10^{-6}$ & 1950 & 1680 \\
$10^{-5}$ & 1670 & 1410 \\
$10^{-4}$ & 1395 & 1125 \\
$10^{-3}$ & 1120 & 855  \\
$10^{-2}$ & 845  & 585  \\
0.7 & 430  & 165  \\
\enddata
\end{deluxetable}

\begin{thebibliography}{}
\bibitem[Avrett(1995)]{avrett95}
Avrett, E. H. 1995, in Infrared Tools for Solar Astrophysics: What's
Next?, ed. J. R. Kuhn, \& M. J. Penn (Singapore: World Scientific), 303
\bibitem[Belluzzi \& Landi Degl'Innocenti(2009)]{lbandeld09}
Belluzzi, L., \& Landi Degl'Innocenti, E. 2009, \aap, 495, 577
\bibitem[Bommier(1997a)]{vb97a}
Bommier, V. 1997a, \aap, 328, 706
\bibitem[Bommier(1997b)]{vb97b}
Bommier, V. 1997b, \aap, 328, 726
\bibitem[Bommier \& Sahal-Br\'echot(1978)]{vbandsb78}
Bommier, V., \& Sahal-Br\'echot, S. 1978, \aap, 69, 57
\bibitem[Bommier \& Stenflo(1999)]{bs99}
Bommier, V., \& Stenflo, J.~O. 1999, \aap, 305, 327
\bibitem[Chandrasekhar(1950)]{chandra50}
Chandrasekhar, S. 1950, Radiative Transfer (Oxford: Clarendon Press)
\bibitem[Domke \& Hubeny(1988)]{dh88}
Domke, H., \& Hubeny, I. 1988, \apj, 334, 527
\bibitem[Dumont et al.(1977)]{dumontetal77}
Dumont, S., Omont, A., Pecker, J.~C., \& Rees, D. E. 1977, \aap, 54, 675
\bibitem[Faurobert(1987)]{mf87}
Faurobert, M. 1987, \aap, 178, 269
\bibitem[Faurobert(1988)]{mf88}
Faurobert, M. 1988, \aap, 194, 268
\bibitem[Faurobert-Scholl(1991)]{mf91}
Faurobert-Scholl, M., 1991, \aap, 246, 469
\bibitem[Faurobert-Scholl(1992)]{mf92}
Faurobert-Scholl, M., 1992, \aap, 258, 521
\bibitem[Faurobert-Scholl(1993)]{mf93}
Faurobert-Scholl, M., 1993, \aap, 268, 765
\bibitem[Fluri et al.(2003)]{flurietal03}
Fluri, D.~M., Nagendra, K.~N., \& Frisch, H. 2003,
\aap, 400, 303
\bibitem[Frisch(1980)]{hf80}
Frisch, H. 1980, \aap, 87, 357
\bibitem[Frisch(1988)]{hf88}
Frisch, H. 1988, in Radiation in Moving Gaseous Media, ed. Y. Chmielewski, \& 
T. Lanz (Switzerland: Geneva Observatory), 337
\bibitem[Frisch(1996)]{hf96}
Frisch, H. 1996, \solphys, 164, 49
\bibitem[Frisch(2007)]{hf07}
Frisch, H. 2007, \aap, 476, 665
\bibitem[Frisch et al.(2009)]{hfetal09}
Frisch, H., Anusha, L.~S., Sampoorna, M., \& Nagendra, K.~N. 2009, \aap, 501, 
335
\bibitem[Gandorfer(2000)]{gan00}
Gandorfer, A. 2000, The Second Solar Spectrum, Vol. I\,: 4625\,\AA\ to
6995\,\AA\ (Zurich\,: vdf Hochschulverlag)
\bibitem[Gandorfer(2002)]{gan02}
Gandorfer, A. 2002, The Second Solar Spectrum, Vol. II\,: 3910\,\AA\ to
4630\,\AA\ (Zurich\,: vdf Hochschulverlag)
\bibitem[Gandorfer(2005)]{gan05}
Gandorfer, A. 2005, The Second Solar Spectrum, Vol. III\,: 3160\,\AA\ to
3915\,\AA\ (Zurich\,: vdf Hochschulverlag)
\bibitem[Hamilton(1947)]{hamilton47}
Hamilton, D.~R. 1947, \apj, 106, 457
\bibitem[Heinzel(1981)]{heinzel81}
Heinzel, P. 1981, \jqsrt, 25, 483
\bibitem[Holzreuter et al.(2005)]{reneetal05}
Holzreuter, R., Fluri, D.~M., \& Stenflo, J.~O. 2005, \aap, 434, 713
\bibitem[Hubeny(1982)]{hubeny82}
Hubeny, I. 1982, \jqsrt, 27, 593
\bibitem[Hubeny(1985a)]{hubeny85a}
Hubeny, I. 1985a, \aap, 145, 463
\bibitem[Hubeny(1985b)]{hubeny85b}
Hubeny, I. 1985b, in Progress in Stellar Spectral Line, 
ed. J.~E. Beckman, \& L. Crivellari (Dordrecht: Reidel), 27
\bibitem[Hubeny \& Cooper(1986)]{hc86}
Hubeny, I., \& Cooper, J. 1986, \apj, 305, 852
\bibitem[Hubeny \& Lites(1995)]{hl95}
Hubeny, I., \& Lites, B.~W. 1995, \apj, 455, 376
\bibitem[Hubeny et al.(1983a)]{hos83a}
Hubeny, I., Oxenius, J., \& Simonneau, E. 1983a, \jqsrt, 29, 477
\bibitem[Hubeny et al.(1983b)]{hos83b}
Hubeny, I., Oxenius, J., \& Simonneau, E. 1983b, \jqsrt, 29, 495
\bibitem[Hummer(1962)]{hum62}
Hummer, D.~G. 1962, \mnras, 125, 21
\bibitem[Kneer(1975)]{kf75}
Kneer, F. 1975, \apj, 200, 367
\bibitem[Landi Degl'Innocenti(1983)]{landi83}
Landi Degl'Innocenti, E. 1983, \solphys, 85, 3
\bibitem[Landi Degl'Innocenti(1984)]{lan84}
Landi Degl'Innocenti, E. 1984, \solphys, 91, 1
\bibitem[Landi Degl'Innocenti(1998)]{landi98}
Landi Degl'Innocenti, E. 1998, \nat, 392, 256
\bibitem[Landi Degl'Innocenti et al.(1997)]{landietal97}
Landi Degl'Innocenti, E., Landi Degl'Innocenti, M., Landolfi, M. 1997, 
in Proc. Forum TH\'EMIS, Science with TH\'EMIS, ed. N. Mein \& S. 
Sahal-Br\'echot (Paris: Obs. Paris-Meudon), 59
\bibitem[Landi Degl'Innocenti \& Landolfi(2004)]{ll04}
Landi Degl'Innocenti, E., \& Landolfi, M. 2004, Polarization in Spectral 
Lines (Dordrecht: Kluwer)
\bibitem[Manso Sainz \& Trujillo Bueno(2003)]{ms-jtb03}
Manso Sainz, R., \& Trujillo Bueno, J. 2003, Phys. Rev. Letters, 91, 111102 
\bibitem[Mihalas(1978)]{mih78}
Mihalas, D. 1978, Stellar Atmosphere (2nd ed.; San Francisco, CA: Freeman)
\bibitem[Nagendra(1994)]{knn94}
Nagendra, K.~N. 1994, \apj, 432, 274
\bibitem[Nagendra(1995)]{knn95}
Nagendra, K.~N. 1995, \mnras, 274, 523
\bibitem[Nagendra et al.(2002)]{knnetal02}
Nagendra, K.~N., Frisch, H., \& Faurobert, M. 2002, \aap, 395, 305
\bibitem[Nagendra et al.(1999)]{knnetal99}
Nagendra, K. N., Paletou, F., Frisch, H., \& Faurobert-Scholl, M. 1999, in 
Solar Polarization, ed. K.~N. Nagendra, \& J.~O. Stenflo (Boston: Kluwer), 127
\bibitem[Nagendra \& Sampoorna(2009)]{knnandms09}
Nagendra, K.~N., \& Sampoorna, M. 2009, in ASP Conf. Ser. 405, Solar
Polarization 5, ed.
S.~V. Berdyugina, K.~N. Nagendra, \& R. Ramelli (San Francisco: ASP), 261
\bibitem[Omont et al.(1972)]{osc72}
Omont, A., Smith, E.~W., \& Cooper, J. 1972, \apj, 175, 185
\bibitem[Omont et al.(1973)]{osc73}
Omont, A., Smith, E.~W., \& Cooper, J. 1973, \apj, 182, 283
\bibitem[Rees \& Saliba(1982)]{rs82}
Rees, D.~E., \& Saliba, G.~J. 1982, \aap, 115, 1  
\bibitem[Saliba(1985)]{gs85}
Saliba, G.~J. 1985, \solphys, 98, 1
\bibitem[Saliba(1986)]{gs86}
Saliba, G. J. 1986, PhD thesis, Univ. of Sidney
\bibitem[Sampoorna et al.(2008b)]{sametal08b}
Sampoorna, M., Nagendra, K. N., \& Frisch, H. 2008b, \jqsrt, 109, 2349
\bibitem[Sampoorna et al.(2007a)]{sametal07a}
Sampoorna, M., Nagendra, K.~N., \& Stenflo, J.~O. 2007a,
\apj, 663, 625
\bibitem[Sampoorna et al.(2007b)]{sametal07b}
Sampoorna, M., Nagendra, K.~N., \& Stenflo, J.~O. 2007b,
\apj, 670, 1485
\bibitem[Sampoorna et al.(2008a)]{sametal08a}
Sampoorna, M., Nagendra, K. N., \& Stenflo, J. O. 2008a, \apj, 679, 889
\bibitem[Sampoorna \& Trujillo Bueno(2010)]{samandjtb10}
Sampoorna, M., \& Trujillo Bueno, J. 2010, \apj, 712, 1331
\bibitem[Stenflo(1994)]{jos94}
Stenflo, J. O. 1994, Solar Magnetic Fields\,: Polarized Radiation Diagnostics 
(Dordrecht: Kluwer)
\bibitem[Stenflo(2004)]{jos04}
Stenflo, J.~O. 2004,  Reviews in Modern Astronomy, 17, 269
\bibitem[Stenflo(2006)]{jos06}
Stenflo, J.~O. 2006, in ASP Conf. Ser. 358, Solar Polarization 4, ed. 
R. Casini, \& B. W. Lites (San Francisco, CA: ASP), 215
\bibitem[Stenflo \&\ Keller(1997)]{josck97}
Stenflo, J.~O., \& Keller, C.~U. 1997, \aap, 321, 927
\bibitem[Stenflo \&\ Stenholm(1976)]{josandslg76}
Stenflo, J.~O., \& Stenholm, L.~G. 1976, \aap, 46, 69
\bibitem[\v St\v ep\'an \& Trujillo Bueno(2010)]{stepan-jtb10}
\v St\v ep\'an, J., \& Trujillo Bueno, J. 2010, \apj, 711, L133
\bibitem[Trujillo Bueno(1999)]{jtb99}
Trujillo Bueno, J. 1999, in Solar Polarization, ed. K.~N. Nagendra, 
\& J.~O. Stenflo (Boston: Kluwer), 73 
\bibitem[Trujillo Bueno(2001)]{jtb01}
Trujillo Bueno, J. 2001, in ASP Conf. Ser. 236, Advanced Solar Polarimetry\,: 
Theory, Observations, and Instrumentation, ed. M. Sigwarth (San Francisco, CA: 
ASP), 161
\bibitem[Trujillo Bueno(2003)]{jtb03}
Trujillo Bueno, J. 2003, in ASP Conf. Ser. 288, 
Stellar Atmosphere Modeling, ed. I. Hubeny, D. Mihalas, \& K.
Werner (San Francisco, CA: ASP), 551
\bibitem[Trujillo Bueno(2009)]{jtb09}
Trujillo Bueno, J. 2009, in ASP Conf. Ser. 405, Solar Polarization 5, ed. 
S. V. Berdyugina, K. N. Nagendra, \& R. Ramelli (San Francisco, CA: ASP), 65
\bibitem[Trujillo Bueno et al.(2004)]{jtb-nature04}
Trujillo Bueno, J., Shchukina, N., \& Asensio Ramos, A. 2004, \nat, 430, 326
\bibitem[Uitenbroek(2003)]{rh03}
Uitenbroek, H. 2003, in ASP Conf. Ser. 288, Stellar atmosphere modeling, 
ed. I. Hubeny, D. Mihalas, \& K. Werner (San Francisco: ASP), 597
\bibitem[Vernazza et al.(1981)]{val81}
Vernazza, J.~E., Avrett, E.~H., \& Loeser, R. 1981, \apjs, 45, 635
\end{thebibliography}
\end{document}